\documentclass[aps,onecolumn,11pt,floatfix,altaffilletter, tightenlines,showpacs,showkeys,notitlepage,nofootinbib]{revtex4-1}
\pdfoutput=1

\usepackage[colorlinks=true,citecolor=blue,linkcolor=blue]{hyperref}
\usepackage[normalem]{ulem}
\usepackage{amsmath,amssymb}
\usepackage{epsfig}
\usepackage{graphicx}
\usepackage{url}
\usepackage{color}
\usepackage{multirow}
\usepackage{placeins}
\usepackage[dvipsnames]{xcolor}
\usepackage{epstopdf}
\usepackage[abs]{overpic}
\usepackage{slashed}

\usepackage{amssymb}
\usepackage{pifont}

\usepackage{tikz}
\usetikzlibrary{trees}
\usetikzlibrary{decorations.pathmorphing}
\usetikzlibrary{decorations.markings}

\definecolor{jblue}  {RGB}{20,50,100}
\definecolor{npurple}  {RGB} {153, 51, 204}
\definecolor{wred}   {RGB}{217,0,56}
\definecolor{white}   {RGB}{255,255,255}

\definecolor{korange}   {RGB}{235, 80,  43}
\definecolor{korange2}   {RGB}{245, 100,  63}
\definecolor{kyelloworange}   {RGB}{255, 210,  110}
\definecolor{kyelloworange2}   {RGB}{240, 170,  90}
\definecolor{kred}   {RGB}{204,  102, 153}
\definecolor{kpurple}   {RGB}{153,  61, 190}
\definecolor{kpurplelight}   {RGB}{213,  161, 230}

\usepackage[capitalize]{cleveref}
\usepackage{subfigure} 
\usepackage{lipsum}
\definecolor{red}{rgb}{1.0, 0, 0}
 \usepackage{gensymb}

\allowdisplaybreaks

\bibpunct{[}{]}{,}{n}{}{,}

\newcommand{\parenbar}[1]{\overset{
            \raisebox{-0.15em}{\scalebox{.4}{\textbf{(}}}
            \raisebox{-0.3em}{{\hspace{.03em}--\hspace{.05em}}}
            \raisebox{-0.15em}{\scalebox{.4}{\textbf{)}}}} {#1}}

\pacs{}
\keywords{}

\begin{document}

\title{
Producing a new Fermion in Coherent Elastic Neutrino-Nucleus Scattering: from Neutrino Mass to Dark Matter 
}

\author{Vedran Brdar}   \email{vbrdar@mpi-hd.mpg.de}
\author{Werner Rodejohann}   \email{werner.rodejohann@mpi-hd.mpg.de}
\author{Xun-Jie Xu} \email{xunjie.xu@mpi-hd.mpg.de}
\affiliation{Max-Planck-Institut f\"ur Kernphysik,
       69117~Heidelberg, Germany}

\begin{abstract}
\noindent
We consider the production of a new MeV-scale fermion in coherent elastic neutrino-nucleus scattering. 
The effect on the measurable nucleon recoil spectrum is calculated. Assuming that the new fermion couples to neutrinos and quarks via a singlet scalar, we set limits on its mass and 
coupling using COHERENT data and also determine the sensitivity of the
CONUS experiment. We investigate the possible connection of the new 
fermion to neutrino mass generation. The possibility of the new fermion being the dark matter particle 
is also studied.

\end{abstract}

\maketitle

\section{Introduction}
\label{sec:intro}
\noindent
Despite being the most elusive Standard Model (SM) particles, neutrinos 
have been detected in a number of charged- and neutral-current processes. The recent measurement \cite{Akimov:2017ade} of coherent elastic neutrino-nucleus scattering (CE$\nu N$S) \cite{Freedman:1973yd,Horowitz:2003cz,Drukier} yields a novel channel where, for the first time, the interaction of low energy neutrinos with nuclei as a whole is probed. This serves not only as a handle to probe SM and nuclear physics parameters, but also as a robust probe of new physics. In particular, light sterile neutrinos \cite{Anderson-nus,Dutta-nus,Kosmos-nus}, non-standard interactions of both quarks and leptons \cite{Dutta:2015vwa,Denton:2018xmq,Lindner:2016wff,Coloma:2017ncl,Liao:2017uzy,Kosmas:2017tsq,Farzan:2018gtr,Abdullah:2018ykz,Billard:2018jnl} as well as neutrino magnetic moments \cite{nu-magnetic1,nu-magnetic2} can be searched for.\\

The basic requirement for the coherent neutrino-nucleus scattering is the smallness
of the momentum transfer. Namely, in case it exceeds the inverse size of the nucleus, one can in principle determine on which nucleon the scattering occurred and this is what breaks the coherence. It is also important that the quantum state of the nucleus does not alter in the scattering because, otherwise, the nuclear excitations in such processes would allow individual nucleons to be tagged which would again directly break the condition for the coherent scattering \cite{Akhmedov:2018wlf}.

On the other hand, the production of new light particles does not a priori violate the coherence as long as the above conditions for nuclei are satisfied. Hence, in this work we explore an interesting new possibility for coherent elastic scattering process, namely $$\nu N \rightarrow \chi N\,.$$ 
Here a light MeV-scale fermion (dubbed $\chi$) is produced from the interaction of the incoming neutrino $\nu$ with a nucleus $N$. 
We are interested, given the lack of evidence for new physics at high energy, 
 in MeV-scale particles as this is the typical 
energy scale of CE$\nu N$S, where naturally the most interesting phenomenology arises. 

Assuming in a minimal setup that the interaction of the new fermion $\chi$ with neutrinos and quarks 
is mediated by a scalar singlet $S$, we derive limits on the 
masses of $\chi$ and $S$ and their coupling to neutrinos and the nucleus. 
Existing and expected data from the running experiments COHERENT \cite{Akimov:2017ade} and CONUS \cite{CONUStalk} is used, 
and the results are  compared to existing terrestrial and astrophysical limits. 
In the near future, other  upcoming experiments including 
$\nu$-cleus \cite{Strauss:2017cuu}, CONNIE \cite{Aguilar-Arevalo:2016khx},
MINER \cite{Agnolet:2016zir}, TEXONO \cite{Wong:2010zzc},  $\nu$GEN \cite{Belov:2015ufh} and Ricochet \cite{Billard:2016giu} will also be able to measure the CE$\nu$NS process.

Any new fermion that couples to light neutrinos needs to be considered regarding its role in the generation of neutrino mass, and we demonstrate that a straightforward extension of the type-I seesaw mechanism can indeed 
generate the observable magnitude of neutrino masses, as well as be testable in CE$\nu N$S. 
Moreover, any new particle beyond the Standard Model is an attractive candidate for dark matter (DM), therefore we investigate 
in such a setup whether $\chi$ can be such a popular MeV-scale DM candidate 
(see e.g.\ Refs.\ \cite{Bertuzzo:2017lwt,Hochberg:2017wce,Dolan:2017xbu,Hufnagel:2017dgo,Dutra:2018gmv,Berlin:2018sjs} for recent studies).  
We find that for the size of the couplings to which CE$\nu N$S experiments are sensitive, 
the DM abundance can match the observed value in case there was an entropy injection episode 
between the QCD phase transition and Big Bang Nucleosynthesis (BBN).\\

The paper is organized as follows. In \cref{sec:production} we derive bounds on the relevant 
couplings and masses within our framework of $\nu N \rightarrow \chi N$ coherent scattering without restricting the discussion to a specific model. We also obtain the corresponding 
recoil spectra of $N$ in case a massive particle $\chi$ is emitted in the final state. 
In \cref{sec:model} we discuss a minimal UV-complete setup in which the MeV-scale $\chi$ is 
related to neutrino mass generation. \cref{sec:DM} is devoted to the assumption that $\chi$ is the 
DM particle, in which we scrutinize its production in the early Universe. 
Finally, in \cref{sec:summary} we conclude. 
\section{Probing MeV-scale Particle in CE\boldmath{$\nu N$}S}
\label{sec:production}
\begin{figure}[t!]
   \centering
    \includegraphics[width=0.35\textwidth]{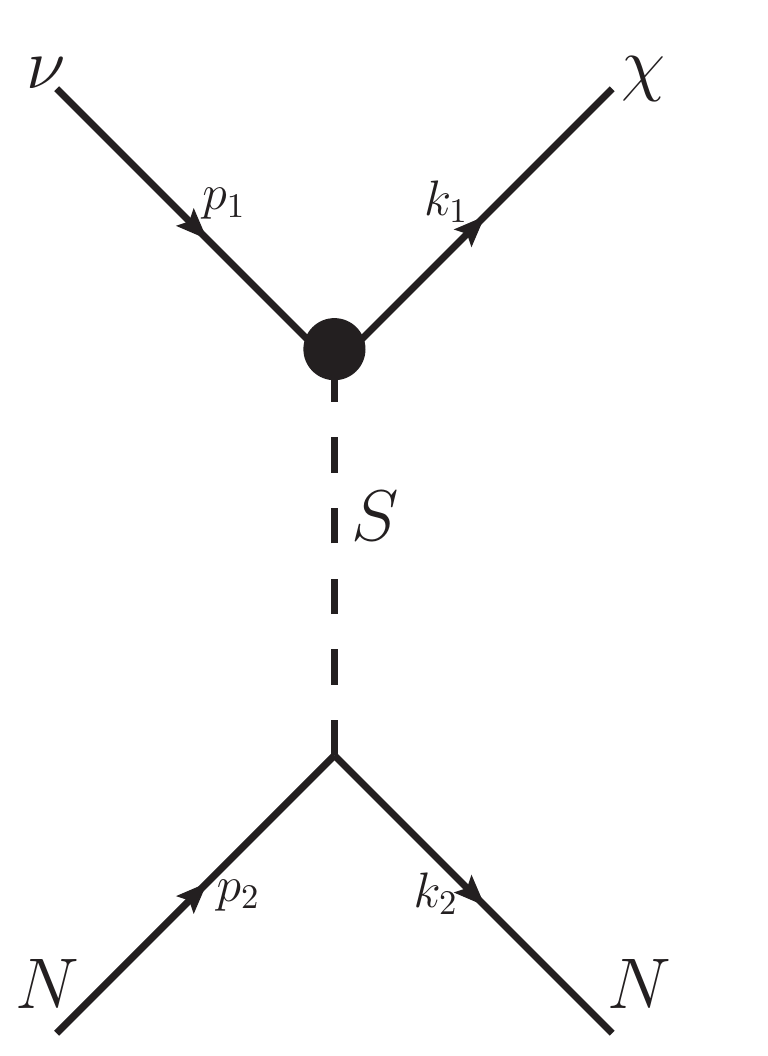}
    \caption{Feynman diagram for $ \nu N\rightarrow  \chi N$ coherent scattering mediated by a scalar $S$. The blob represents the (possibly effective, see Section \ref{subsec:model}) $\nu$-$S$-$\chi$ vertex. }
	\label{fig1}
\end{figure}

\noindent
In this section we investigate the phenomenological aspects of  $\nu N \rightarrow \chi N$ coherent scattering by assuming only the following interaction
\begin{align}
\mathcal{L} \supset y_{\chi} \bar{\chi} S \nu + y_{N}\bar{N} S N\,,
\label{eq:eft-lagrangian}
\end{align}
where $y_\chi$ and $y_N$ parametrize 
the strength of the Yukawa interaction of a mediator particle $S$ with 
$\nu$-$\chi$ and the nucleus, respectively.
In principle, the mediator for a $2\to 2$ process with fermions on the external legs
can be a scalar or vector boson; we will consider scalar mediators here, though the discussion in Section \ref{sec:2a} is independent on this. 
Furthermore, we do not require significant mixing between active neutrinos and $\chi$ 
for coherent scattering, and hence the exchange of SM gauge bosons is suppressed.  
Model building options for generating interactions of a scalar singlet $S$ with quarks, and 
hence eventually nuclei, are presented for instance in Ref.\ \cite{Farzan:2018gtr}. 
The process under our consideration is shown in a diagrammatic form in \cref{fig1}.

\subsection{\label{sec:2a}Prerequisites for obtaining the cross sections}
\noindent
Due to the mass of $\chi$, the process $\nu N\rightarrow\chi N$
has different kinematics than CE$\nu N$S. Hence, as a starting point,
we derive some relations for the kinematics of this process 
that will be used throughout the paper. 
The notation of various quantities is given as follows: 
\begin{itemize}
 \setlength\itemsep{0.5em}
\item[$-$] $p_{1}^{\mu}$ and $p_{2}^{\mu}$ denote the initial 4-momenta of the
neutrino and the nucleus, respectively;
\item[$-$] $k_{1}^{\mu}$ and $k_{2}^{\mu}$ denote the final 4-momenta of $\chi$ and the nucleus, respectively;
\item[$-$] $q^{\mu}\equiv k_{2}^{\mu}-p_{2}^{\mu}=p_{1}^{\mu}-k_{1}^{\mu}$ denotes
the momentum transfer in the scattering process;
\item[$-$] $M$, $m_{\chi}$, and $m_{S}$ denote the masses of the nucleus,
 $\chi$, and the mediator $S$, respectively;
\item[$-$] $E_{\nu}$ is the energy of the incoming neutrino;
\item[$-$] $T$ denotes the recoil energy of the nucleus, and $p=\sqrt{(M+T)^{2}-M^{2}}$
is the recoil momentum;
\item[$-$] $\theta$ is the angle of the outgoing nucleus with respect
to the incoming neutrino; i.e.\ the angle between $\boldsymbol{k}_{2}$
and $\boldsymbol{p}_{1}$. 
\end{itemize}

Using the above notations, we can explicitly express the 4-momenta:
\begin{subequations}
\begin{alignat}{2}
p_{1}^{\mu}&=(E_{\nu},\ E_{\nu},\ 0,\ 0)\,,\label{eq:cohdm-2} \\
p_{2}^{\mu}&=(M,\ 0,\ 0,\ 0)\,, &  k_{2}^{\mu}&=(M+T,\ p\cos\theta,\ p\sin\theta,\ 0)\,,\label{eq:cohdm-3}\\
q^{\mu}&\equiv k_{2}^{\mu}-p_{2}^{\mu}=(T,\ p\cos\theta,\ p\sin\theta,\ 0)\,.\label{eq:cohdm-4}
\end{alignat}
\end{subequations}
When computing the cross section, scalar products of the external
momenta (e.g.\ $p_{1}\cdot p_{2}$, $p_{1}\cdot k_{1}$, $p_{2}\cdot k_{2}$, etc.)
will be used. All scalar products of $p_{1}^{\mu}$,
$p_{2}^{\mu}$ and $q^{\mu}$ ($k_{1}^{\mu}$ and $k_{2}^{\mu}$ can
be expressed in terms of these three 4-momenta) read:
\begin{align}
&p_{1}^{2} =0\,, &  &p_{2}^{2} =M^{2}\,, &  &q^{2}=-2MT\,,\nonumber \\
&p_{1} \cdot p_{2}=ME_{\nu}\,, &  &p_{1}\cdot q =-MT-m_{\chi}^{2}/2\,, &  &p_{2} \cdot q =MT\,.
\label{eq:cohdm-5}
\end{align}
We obtained $q^{2}$ by squaring both sides of $q^{\mu}\equiv k_{2}^{\mu}-p_{2}^{\mu}$
and using Eq.~(\ref{eq:cohdm-3}):
\[
q^{2}=k_{2}^{2}+p_{2}^{2}-2k_{2}.p_{2}=2M^{2}-2M(M+T)=-2MT.
\]
Applying the same to $p_{2}^{\mu}+q^{\mu}=k_{2}^{\mu}$ and
$p_{1}^{\mu}-q^{\mu}=k_{1}^{\mu}$ and using $q^{2}=-2MT$, we obtained $p_{2} \cdot q$ and $p_{1} \cdot q$ 
given in Eq.~(\ref{eq:cohdm-5}).

\begin{figure}
\centering
\includegraphics[width=8cm]{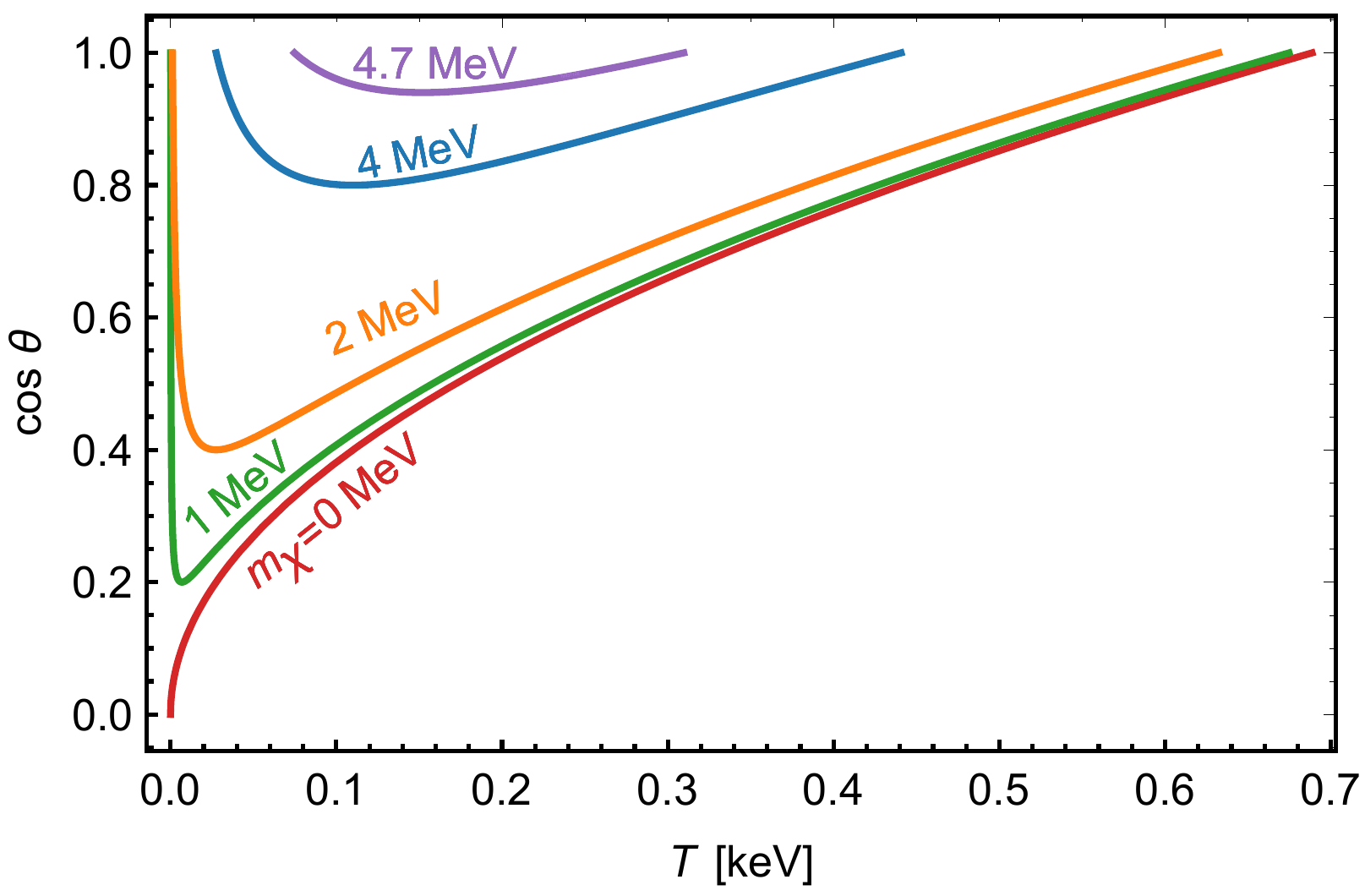}
\caption{\label{fig:Relation}Relation between $\cos\theta$ and $T$. This figure
is produced according to Eq.~(\ref{eq:cohdm-8}) with $M=72.6$ 
GeV (i.e.\ a Germanium detector) and $E_{\nu}=5$ MeV. The case $m_\chi=0$ has the same kinematics  as the standard 
coherent elastic neutrino-nucleon scattering.}
\end{figure}

One can also use the explicit forms of $p_{1}^{\mu}$,
$p_{2}^{\mu}$ and $q^{\mu}$ in Eqs.~(\ref{eq:cohdm-2})-(\ref{eq:cohdm-4})
to compute these scalar products directly, e.g., 
\begin{equation}
p_{1} \cdot q=E_{\nu}T-E_{\nu}\sqrt{(M+T)^{2}-M^{2}}\cos\theta\,.\label{eq:cohdm-7}
\end{equation}
We can compare this result with Eq.~(\ref{eq:cohdm-5})
and obtain 
\begin{equation}
\cos\theta=\frac{E_{\nu}T+MT+m_{\chi}^{2}/2}{E_{\nu}\sqrt{(M+T)^{2}-M^{2}}}\,,\label{eq:cohdm-8}
\end{equation}
which reveals the relation between $\theta$ and $T$. In
Fig.~\ref{fig:Relation} we plot the relation for some specific values of $m_{\chi}$ in order to illustrate
how $\cos\theta$ varies with $T$. Typically (for nonzero $m_{\chi}$)
$\cos\theta$ as a function of $T$ has a minimum corresponding to
the maximal scattering angle $\theta_{{\rm {\rm max}}}$.

By solving $d\cos\theta/dT=0$ we obtain
\begin{align}
\cos\theta_{{\rm {\rm max}}}&=\frac{m_{\chi}\sqrt{4M\left(E_{\nu}+M\right)-m_{\chi}^{2}}}{2ME_{\nu}}\,, &
 T_{\theta{\rm max}}&=\frac{Mm_{\chi}^{2}}{2ME_{\nu}-m_{\chi}^{2}+2M^{2}}\,.\label{eq:cohdm-10}
\end{align}
For $T>T_{\theta{\rm max}}$ ($T<T_{\theta{\rm max}}$), $\cos\theta$
increases (decreases) with $T$. Therefore, $\cos\theta$ should be
in the range
\begin{equation}
\cos\theta_{{\rm {\rm max}}}\leq\cos\theta\leq1\,,
\label{eq:cohdm-11}
\end{equation}
and due to the upper bound, $T$ can reach values in  the range
\begin{equation}
T_{\min}\leq T\leq T_{{\rm max}}\,,\label{eq:cohdm-1}
\end{equation}
where $T_{\min}$ and $T_{\max}$ are determined by setting the left-hand
side of Eq.~(\ref{eq:cohdm-8}) to $1$ and solving the equation with respect
to $T$. The solutions are
\begin{equation}
T_{{\rm min/max}}=\frac{2ME_{\nu}^{2}-m_{\chi}^{2}\left(E_{\nu}+M\right)\mp E_{\nu}\sqrt{4M^{2}E_{\nu}^{2}-4Mm_{\chi}^{2}\left(E_{\nu}+M\right)+m_{\chi}^{4}}}{2M\left(2E_{\nu}+M\right)}\,.\label{eq:cohdm-12}
\end{equation}
One can check that \cref{eq:cohdm-12}
has the following massless limit
\begin{equation}
\lim_{m_{\chi}\rightarrow0}(T_{{\rm min}},\  T_{{\rm max}})=\left(0,\  \frac{2E_{\nu}^{2}}{M+2E_{\nu}}\right),\label{eq:cohdm-13}
\end{equation}
which is consistent with the standard results of coherent elastic neutrino scattering.

Another important quantity is the minimal neutrino energy $E_{\nu}^{\min}$
necessary to create a massive particle $\chi$: 
\begin{equation}
E_{\nu}^{\min}=m_{\chi}+\frac{m_{\chi}^{2}}{2M}\,,\label{eq:cohdm-23}
\end{equation}
which is obtained by solving $T_{{\rm min}}=T_{{\rm max}}$. If
$E_{\nu}$ is lower than $E_{\nu}^{\min}$, $\chi$ cannot be produced
in the scattering. In the limit when $\chi$ can just be produced, 
we have 
\begin{equation}
\lim_{E_{\nu}\rightarrow E_{\nu}^{\min}}T_{{\rm min}}=\lim_{E_{\nu}\rightarrow E_{\nu}^{\min}}T_{{\rm max}}=\frac{m_{\chi}^{2}}{2(M+m_{\chi})}\,.\label{eq:cohdm-24}
\end{equation}

An interesting difference between the cases of massive and massless $\chi$ occurs at $T_{{\rm min}}$. From Eq.~(\ref{eq:cohdm-8}) one can obtain 
\begin{equation}
T\rightarrow T_{\min}\ \ \Longrightarrow\ \ \cos\theta\rightarrow
\begin{cases}
1, & {\rm for}\ m_{\chi}\neq 0\,,\\
0, & {\rm for}\ m_{\chi}= 0\,,
\end{cases}\label{eq:cohdm-14}
\end{equation}
which implies that in the minimal recoil limit for massive $\chi$ the nucleus after scattering
moves along the same direction as the incoming neutrino ($\theta=0$), while
for massless $\chi$ it moves in the perpendicular direction ($\theta=90^{\circ}$).

We would like to clarify here that we are discussing $T$ approaching $T_{\min}$ instead of being exactly equal to $T_{\min}$, because for $m_{\chi}=0$, according to Eq.~(\ref{eq:cohdm-13}),  $T_{\min}$ is exactly zero. If $T=T_{\min}=0$, strictly speaking, $\cos\theta$ is not well defined because it implies that the nucleus after scattering stays at rest. 
If $T$ is approaching $T_{\min}$ but remains nonzero, then $\cos\theta$ indeed is very close to zero for $m_{\chi}=0$. 
For $T$ fixed at a very small but nonzero value, when $m_{\chi}$ increases from zero to nonzero values,  $\cos\theta$ will rise steeply (depending on the smallness of $T$) to 1\,---\,as shown in Fig.~\ref{fig:Relation}.  Therefore there is no inconsistency in the minimal recoil limit of $m_{\chi}=0$ and $m_{\chi}\neq 0$. 
Although the $T$\,-\,$\cos\theta$ relation in the minimal recoil limit is very sensitive to small $m_{\chi}$, experimentally it is difficult to observe this behavior due to rather small recoil energies.

\subsection{Cross sections}
\noindent
The exchanged scalar $S$ is generally assumed to be massive  with its mass denoted
by $m_{S}$. We evaluate the cross section without assuming $m_{S}^{2}\ll q^{2}$ or
$m_{S}^{2}\gg q^{2}$. The heavy/light mass limits will be discussed
below.

From the Feynman diagram in \cref{fig1} and the relevant Lagrangian (\ref{eq:eft-lagrangian}), one can straightforwardly write down the scattering
amplitudes for (anti)neutrino initial state
\begin{align}
i{\cal M}(\nu N\rightarrow\chi N)=\bar{u}^{s'}(k_{1})\left(iy_{\chi}\right)P_{L}u^{s}(p_{1})\frac{-i}{q^{2}-m^{2}}\bar{u}^{r'}(k_{2})\left(iy_{N}\right)u^{r}(p_{2})\,,\label{eq:cohdm-17}
\end{align}
\begin{align}
i{\cal M}(\bar{\nu}N\rightarrow\bar{\chi}N)=\bar{v}^{s}(p_{1})P_{R}\left(iy_{\chi}\right)v^{s'}(k_{1})\frac{-i}{q^{2}-m^{2}}\bar{u}^{r'}(k_{2})\left(iy_{N}\right)u^{r}(p_{2})\,,\label{eq:cohdm-18}
\end{align}
where spinor superscripts denote spin and  we have inserted the left-/right-handed
projectors $P_{L}\equiv(1-\gamma^{5})/2$ and $P_{R}\equiv(1+\gamma^{5})/2$
since the neutrino sources can only emit left-handed neutrinos or
right-handed antineutrinos. Using FeynCalc \cite{Mertig:1990an,Shtabovenko:2016sxi} we compute
$|i{\cal M}|^{2}$ for both cases. The result is identical for both
neutrino and antineutrino scattering, namely 
\begin{equation}
|i{\cal M}|^{2}=\frac{8E_{\nu}^{2}M^{2}y^{4}}{\left(2MT+m_{S}^{2}\right)^{2}}\,K\,,\label{eq:cohdm-19}
\end{equation}
with the combined coupling constant  
\begin{equation}
y^{4}=y_{\chi}^{2}y_{N}^{2}\,.\label{eq:cohdm-22}
\end{equation}
The dimensionless quantity $K$ is typically ${\cal O}(1)$ and reads 
\begin{equation}
K=\left(1+\frac{T}{2M}\right)\left(\frac{MT}{E_{\nu}^{2}}+\frac{m_{\chi}^{2}}{2E_{\nu}^{2}}\right).\label{eq:cohdm-21}
\end{equation}
We will in what follows set limits using experiments with different nuclear targets. 
To reduce the dependence of the limits on the type of the nucleus we define
\begin{equation}
\bar{y}\equiv\frac{y}{\sqrt{A}}\,,\label{eq:cohdm-25}
\end{equation}
where $A$ is the nucleon number (sum of neutron and proton numbers).
Since $\sqrt{A}$ has been factored out, $\bar{y}$ has little
dependence on the type of nuclei. For example, for Ge and CsI detectors
 we obtain\footnote{See Sec.~II.C of Ref.~\cite{Farzan:2018gtr}, where the conversion 
from $y_{n}$ and $y_{p}$ to more fundamental quark Yukawa couplings 
is discussed. }
\begin{equation}
\bar{y}\approx\begin{cases}
\sqrt{\big|(0.56y_{n}+0.44y_{p})y_{\chi}\big|} & ({\rm for\ Ge\ target})\,,\\
\sqrt{\big|(0.58y_{n}+0.42y_{p})y_{\chi}\big|} & ({\rm for\ CsI\ target})\,,
\end{cases}\label{eq:cohdm-26}
\end{equation}
where the Yukawa
couplings of the scalar $S$ to neutrons and protons are denoted with $y_{n}$ and
$y_{p}$ respectively. Clearly, $\bar{y}$ for Ge (employed at the CONUS experiment) is approximately the same as $\bar{y}$ for CsI (currently employed at the COHERENT experiment). 

The differential cross section, according to Eq.~(\ref{eq:cohdm-19}),
reads
\begin{equation}
\frac{d\sigma}{dT}=\frac{|i{\cal M}|^{2}}{32\pi ME_{\nu}^{2}}=\frac{M\bar{y}^{4}}{4\pi A^{2}\left(2MT+m_{S}^{2}\right)^{2}}K\,.\label{eq:cohdm-20}
\end{equation}
One can straightforwardly check that in the limit  
$m_{\chi}^{2}\rightarrow0$ the result in Eq.~(\ref{eq:cohdm-20}) is consistent with the standard 
cross section of elastic neutrino scattering \cite{Akimov:2017ade}.

\subsection{Signals and constraints}
\noindent
Now let us study the signal of our new fermion $\chi$ in CE$\nu N$S 
experiments. We will focus on two experiments, namely COHERENT \cite{Akimov:2017ade}
and CONUS \cite{CONUStalk}. For the former, we will present the limits on the relevant parameters in $\nu N \to \chi N$  scattering based on the recent data release, whereas for the latter experiment we obtain sensitivities. 

The COHERENT experiment is based on neutrino emission from the
Spallation Neutron Source at Oak Ridge National Laboratory.
A crystal scintillator detector with 14.6 kg CsI was used in its recent
measurement of CE$\nu N$S and the SM signal has been observed with
6.7$\sigma$ confidence. The neutrinos are produced via $\pi^{+}$
decay ($\pi^{+}\rightarrow\mu^{+}+\nu_{\mu}$) and subsequently $\mu^{+}$
decay ($\mu^{+}\rightarrow e^{+}+\bar{\nu}_{\mu}+\nu_{e}$).
In this experiment, both $\pi^{+}$ and $\mu^{+}$ approximately decay
at rest, which allows us to obtain the analytical expressions for neutrino spectra
\cite{Coloma:2017egw}
\begin{align}
\phi_{\nu_{\mu}}(E_{\nu}) & =\phi_{0}\,\delta\left(E_{\nu}-E_{\nu0}\right),\ \ \ \ \ \ {\rm with}\ E_{\nu0}=\frac{m_{\pi}^{2}-m_{\mu}^{2}}{2m_{\pi}}\approx29.8\ {\rm MeV},\label{eq:cohdm-27}\\
\phi_{\bar{\nu}_{\mu}}(E_{\nu}) & =\phi_{0}\frac{64E_{\nu}^{2}}{m_{\mu}^{3}}\left(\frac{3}{4}-\frac{E_{\nu}}{m_{\mu}}\right),\ \ \ \left(E_{\nu}<m_{\mu}/2\right),\label{eq:cohdm-30}\\
\phi_{\nu_{e}}(E_{\nu}) & =\phi_{0}\frac{192E_{\nu}^{2}}{m_{\mu}^{3}}\left(\frac{1}{2}-\frac{E_{\nu}}{m_{\mu}}\right),\ \ \ \left(E_{\nu}<m_{\mu}/2\right),\label{eq:cohdm-31}
\end{align}
which contains a monochromatic component $\phi_{\nu_{\mu}}$ (i.e.\ all $\nu_{\mu}$  have the same energy $E_{\nu0}$).\\

The CONUS experiment measures CE$\nu N$S of reactor neutrinos ($\bar{\nu}_{e}$)
from a 3.9 GW nuclear power plant in Brokdorf, Germany.
The detector is a Germanium semiconductor containing 4 kg of
natural Ge ($A=72.6$ in average), which is set at a distance of $17$ meters from the
reactor. To compute the event rates we adopt the reactor neutrino
flux computed in \cite{Huber:2011wv,Mueller:2011nm} and normalize the total
flux to $2.5\times10^{13}\,{\rm s}^{-1}\, {\rm cm}^{-2}$. CONUS data taking has started in 
April 2018 and recently a preliminary 2.4$\sigma$ statistical significance for observing  
the process was announced \cite{CONUStalk}. 

The event numbers in both experiments can be computed in the following
way: in the $i$-th recoil energy bin ($T_{i}<T<T_{i}+\Delta T$),
the total event number\footnote{In COHERENT experiment there are 3 neutrino species (see 
\cref{eq:cohdm-27,eq:cohdm-30,eq:cohdm-31}). In that case the total event rate is obtained by summing individual contributions from the three species.}  $N_{i}$ consists of the SM contribution $N_{i}^{{\rm SM}}$
and the new physics contribution $N_{i}^{{\rm new}}$, i.e.  
\begin{equation}
N_{i}=N_{i}^{{\rm SM}}+N_{i}^{{\rm new}}\,,\label{eq:cohdm-28}
\end{equation}
which are computed by
\begin{equation}
N_{i}^{{\rm SM}}=\Delta t\,N_{{\rm nucleus}}\,\int_{T_{i}}^{T_{i}+\Delta T}dT\int dE_{\nu}\,\phi(E_{\nu})\times\frac{d\sigma^{{\rm SM}}}{dT}\left(T,\thinspace E_{\nu}\right)\times\Theta^{{\rm SM}}\left(T,\thinspace E_{\nu}\right),\label{eq:cohdm-29}
\end{equation}
\begin{equation}
N_{i}^{{\rm new}}=\Delta t\,N_{{\rm nucleus}}\,\int_{T_{i}}^{T_{i}+\Delta T}dT\int dE_{\nu}\,\phi(E_{\nu})\times\frac{d\sigma^{{\rm new}}}{dT}\left(T,\thinspace E_{\nu}\right)\times\Theta^{{\rm {\rm new}}}\left(T,\thinspace E_{\nu}\right),\label{eq:cohdm-32}
\end{equation}
where 
\begin{align}
&\Theta^{{\rm SM}}\left(T,\thinspace E_{\nu}\right)\equiv\begin{cases}
1 & 0<T<T_{\max}^{{\rm SM}}(E_{\nu})\,,\\ 
0 & {\rm otherwise}\,,
\end{cases}\nonumber \\  &\Theta^{{\rm new}}\left(T,\thinspace E_{\nu}\right)\equiv\begin{cases}
1 & T_{\min}(E_{\nu})<T<T_{\max}(E_{\nu})\,,\\
0 & {\rm otherwise}\,.
\end{cases}\label{eq:cohdm-33}
\end{align}
Here $\phi$ is the neutrino spectrum, $N_{{\rm nucleus}}$ is the number of nuclei in the detector and $t$ is the data taking period. The explicit expressions of $T_{\max}^{{\rm SM}}$, $T_{\min}$ and
$T_{\max}$ are given in \cref{eq:cohdm-13,eq:cohdm-12}. We note that we have included a 
form factor $F(q^{2})$ in the cross section for the COHERENT experiment, where we take the
 parametrization given in Ref.\ \cite{Kerman:2016jqp}, see Fig.~1(a) therein. 
For the CsI detectors used in COHERENT, since the atomic number of Xe (54) is between Cs (55) 
and I (53), it is a good approximation to use the Xe form factor for both Cs and I. 
For reactor neutrinos, we can set  $F(q^{2}) = 1$ due to the low recoil energy. 

\begin{figure}
\centering

\includegraphics[width=8.5cm,height=6cm]{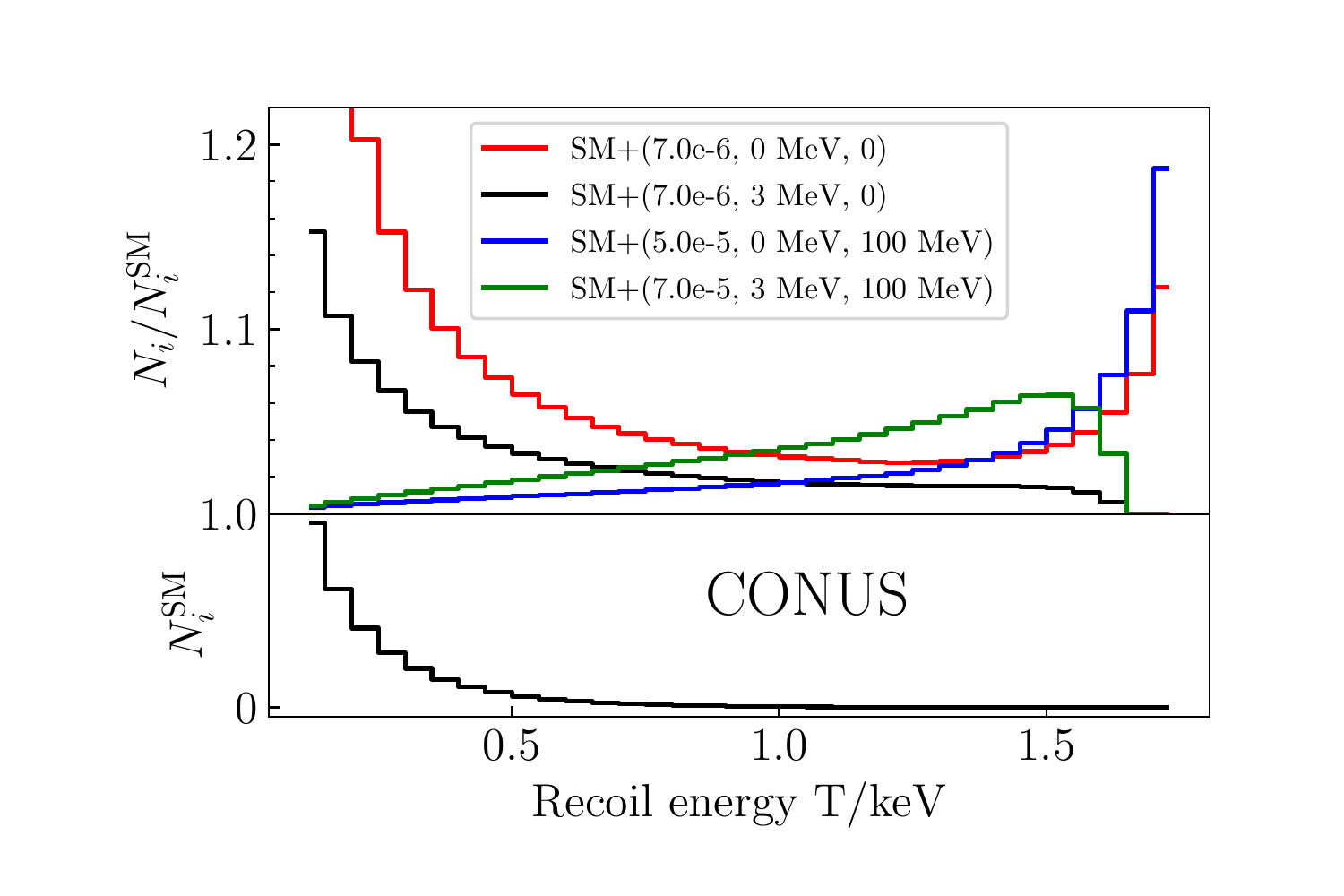}\includegraphics[width=8.5cm,height=6cm]{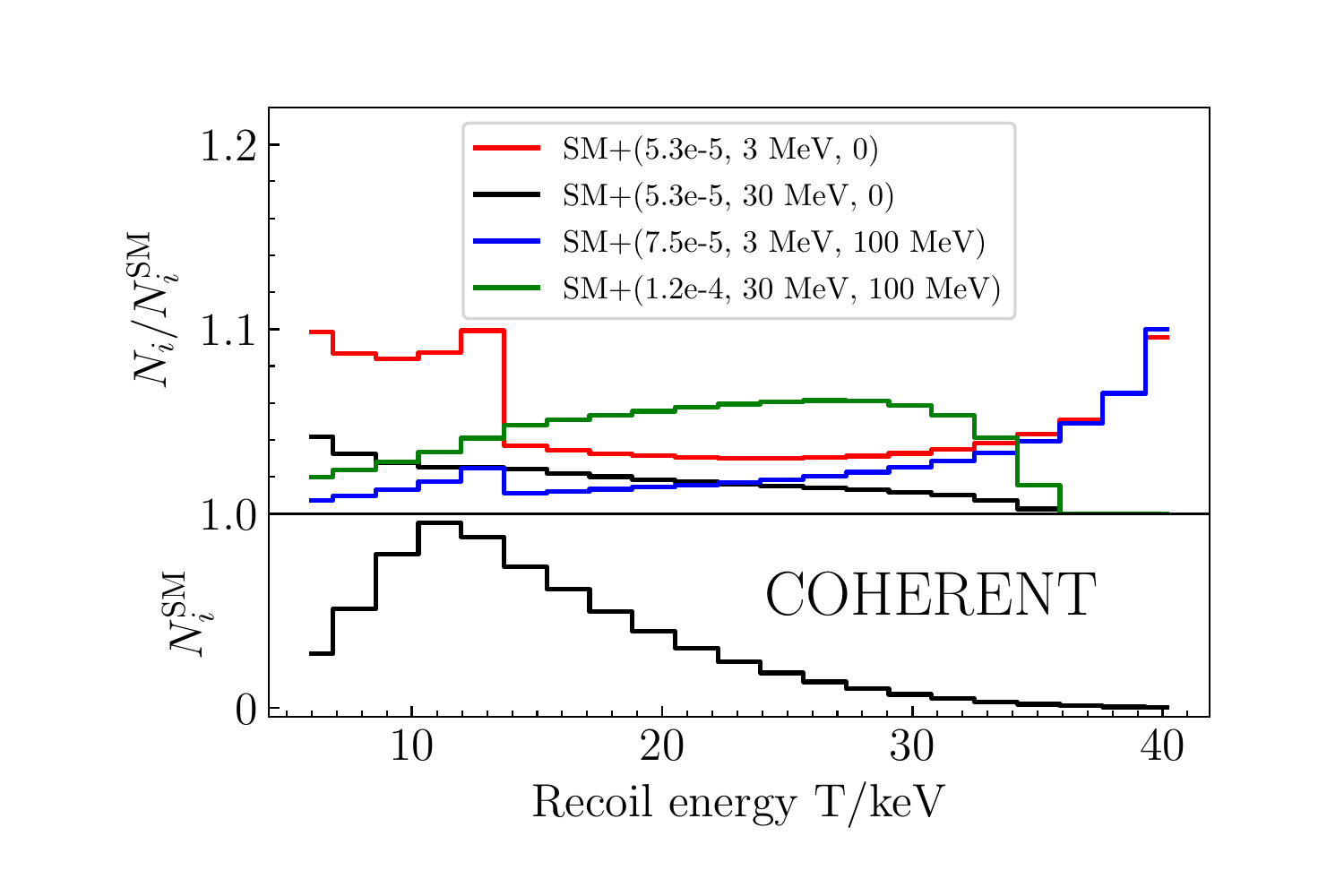}

\caption{\label{fig:signal}Event distributions and spectrum distortions due
to the new  channel $(\nu N \to \chi N)$ in CONUS (left) and COHERENT (right). $N_{i}^{{\rm SM}}$
and $N_{i}$ are the event numbers for  SM-only and SM plus new physics,
respectively. In the upper panels the ratio of the total and SM-only event rate is shown with the corresponding $(\bar{y},\ m_{\chi},\ m_{S})$  indicated in the parentheses. In the lower panels we indicate the recoil spectrum of the SM process in arbitrary units.}
\end{figure}

Using the above equations, we can compute the event numbers and study
the signal of new physics in these two experiments. In Fig.~\ref{fig:signal},
we present the event distributions for several choices of $(\bar{y},\ m_{\chi},\ m_{S})$ 
parameters together with the ratio of $N_{i}/N_{i}^{{\rm SM}}$
for both CONUS (left) and COHERENT (right). We selectively choose
several values for $m_{\chi}$ (0 MeV and 3 MeV for CONUS; 3 MeV and
30 MeV for COHERENT) to illustrate the effect of $m_{\chi}$ on CE$\nu N$S.
Light and heavy mediator cases have been illustrated
by considering both $m_{S}=0$ MeV and $m_{S}=100$ MeV. The
kinks of the red and blue curves appearing in the right panel at $T\approx14$
keV are caused by the monochromatic $\phi_{\nu_{\mu}}$ in COHERENT. The green
and black curves correspond to $m_{\chi}=30$ MeV. Since the monochromatic $\nu_{\mu}$
neutrinos of $29.8$ MeV energy do not have sufficient energy to produce
$\chi$ there are no similar kinks in these two curves. 

\begin{figure}
\centering

\includegraphics[width=7.5cm]{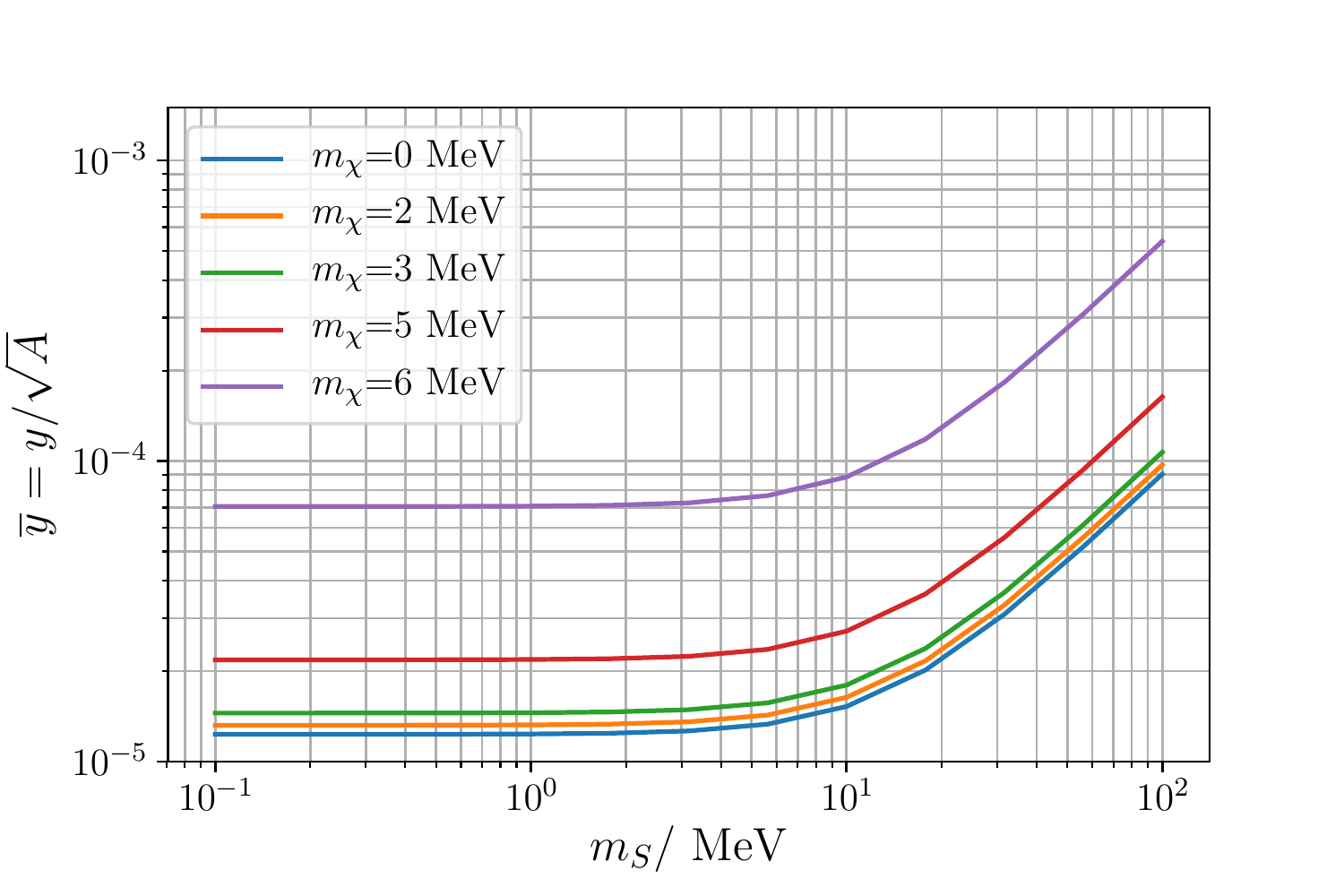}\includegraphics[width=7.5cm]{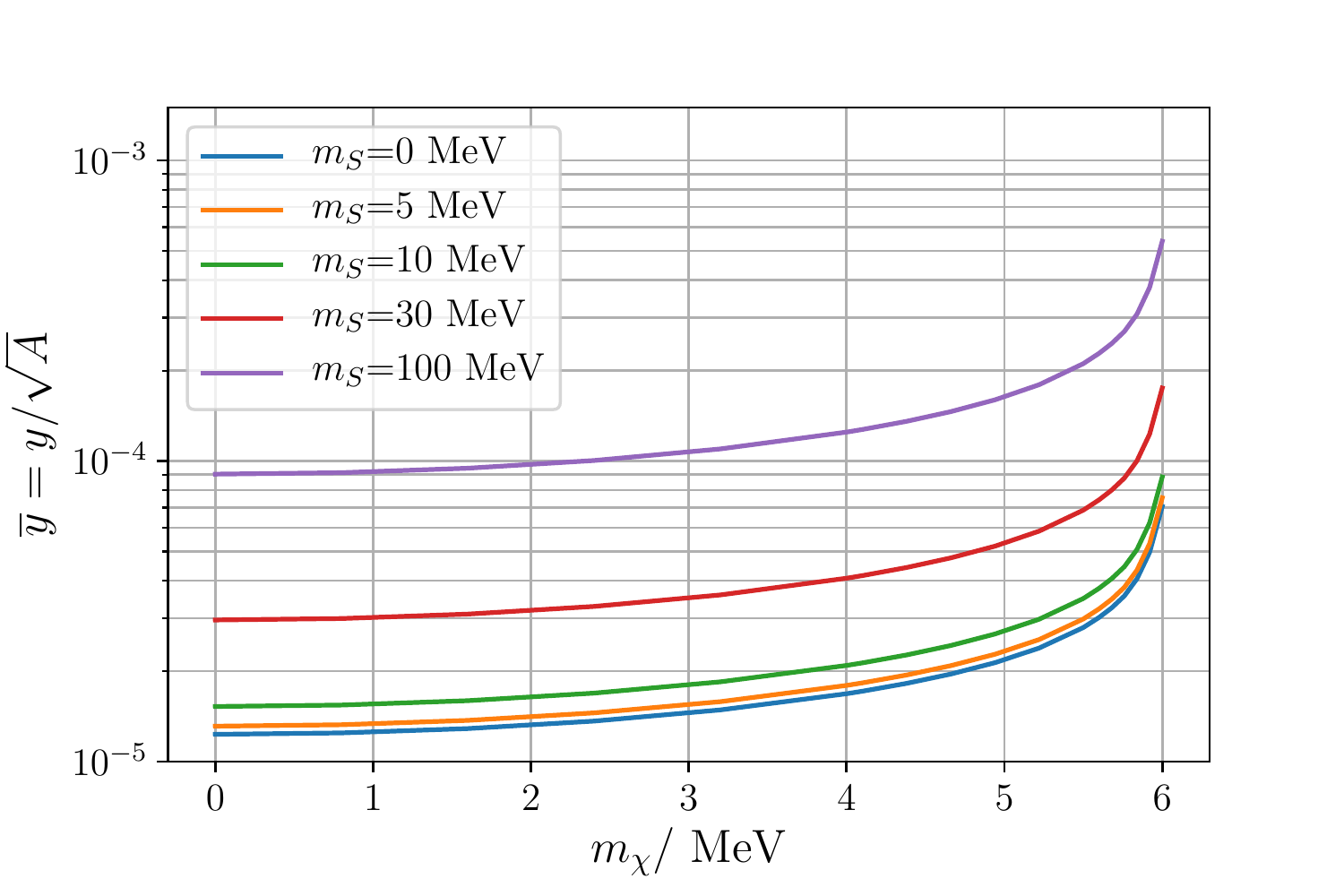}

\includegraphics[width=7.5cm]{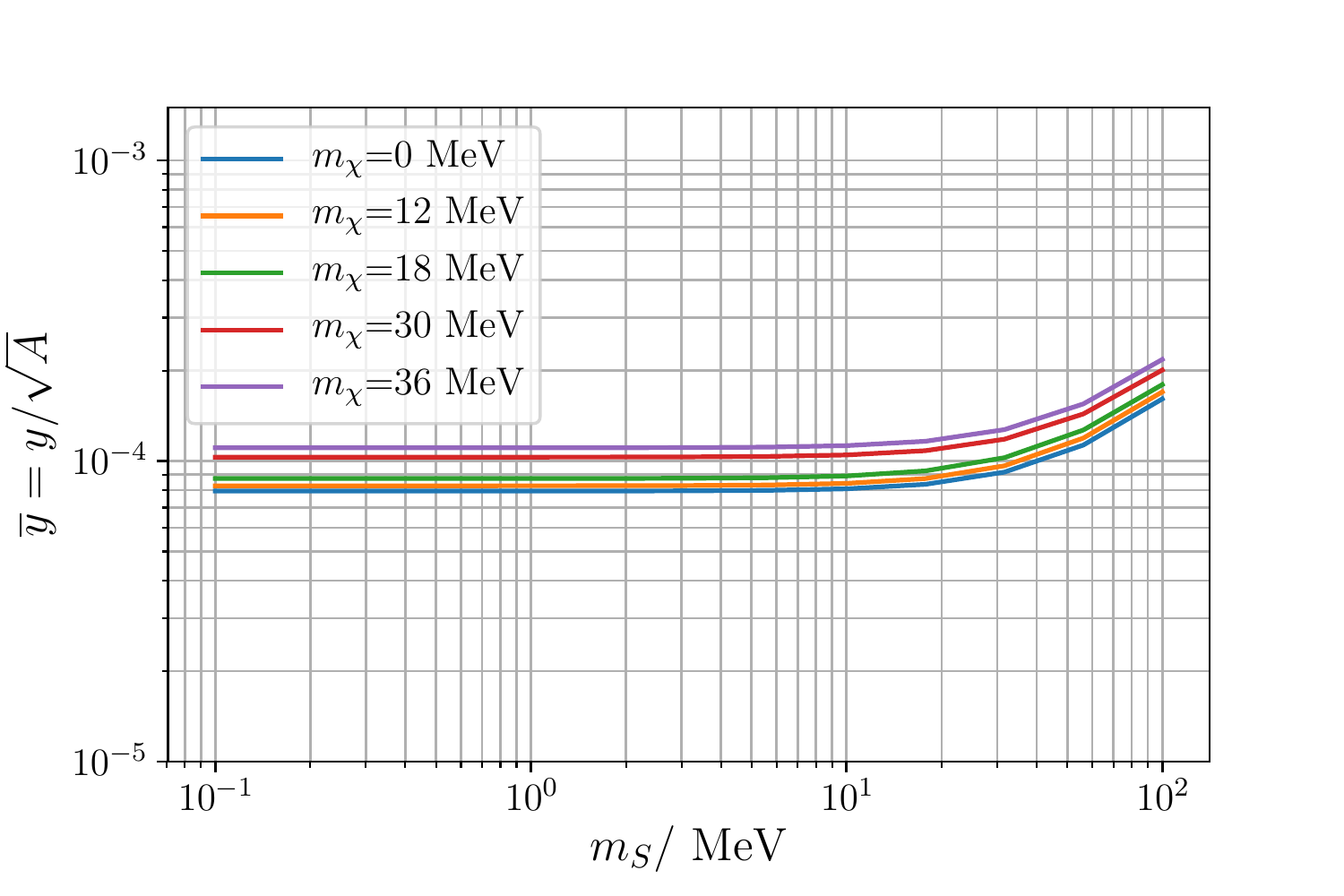}\includegraphics[width=7.5cm]{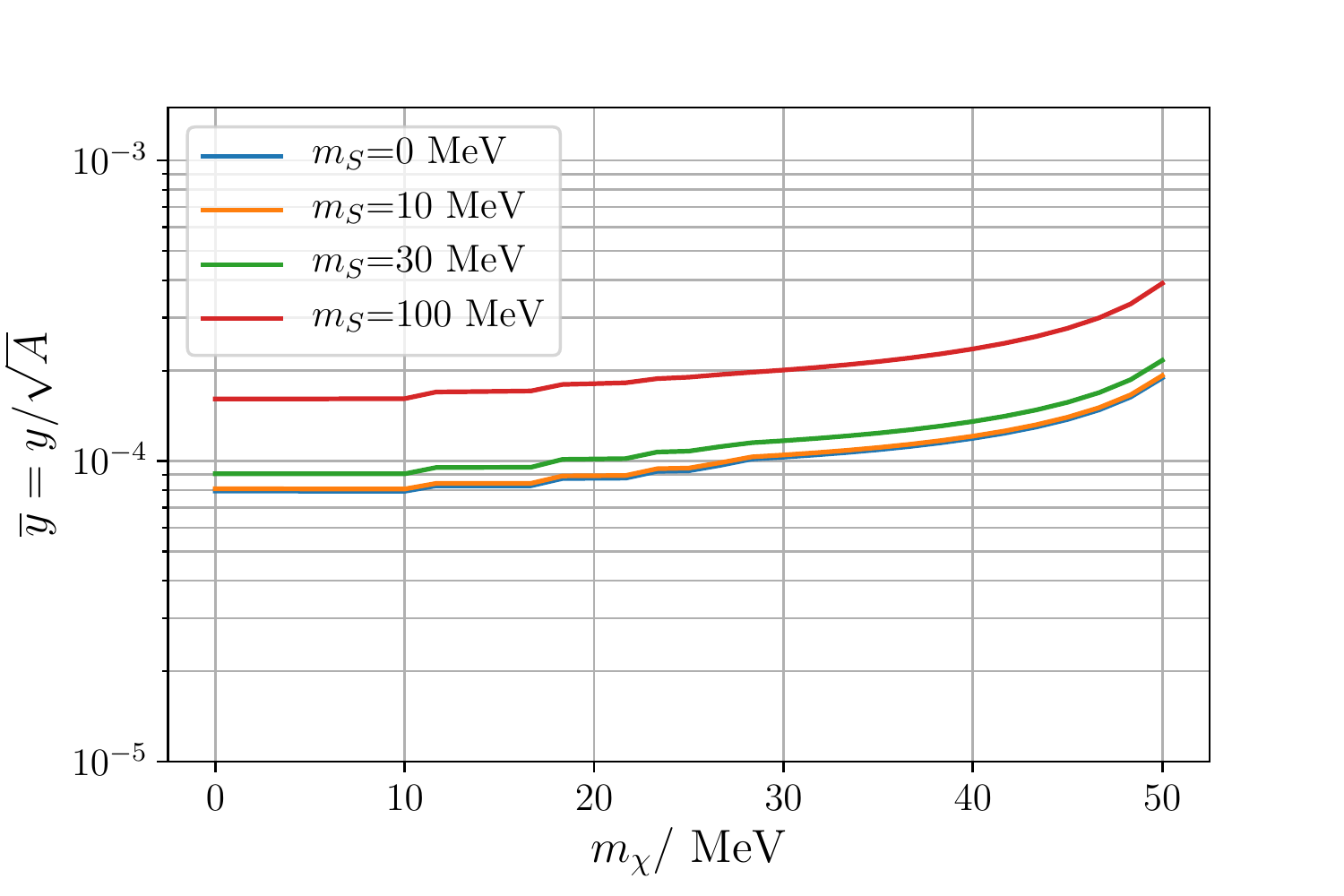}

\caption{\label{fig:constraints}Constraints (with $90\%$ C.L.) of CONUS (upper
panels) and COHERENT (lower panels) in $\bar{y}-m_{S}$ (left
panels) and $\bar{y}-m_{\chi}$ (right panels) parameter space.}
\end{figure}

By comparing $N_{i}$ with the observed event numbers we can obtain
the constraints on the $\chi$ coupling to neutrinos and nuclei. For COHERENT, the observed event
numbers have been published in Ref. \cite{Akimov:2017ade} which can be
used directly in our data fitting procedure. The recoil threshold in COHERENT
is controlled by the signal acceptance fraction (see Fig.~S9 of Ref.~\cite{Akimov:2017ade})
which drops down quickly when  the number of photoelectrons $n_{{\rm PE}}$
($n_{{\rm PE}}\approx1.17\, T/{\rm keV}$) is less than $20$, and approximately
vanishes when $n_{{\rm PE}}<5$. Therefore, in fitting the COHERENT
data we import the signal acceptance fraction directly instead of setting a distinct
threshold. The systematic and statistical uncertainties have been
combined and provided in Fig.~3 of Ref.~\cite{Akimov:2017ade},
and are employed directly in our data fitting.

The CONUS data has not been published, and hence we assume that their findings
will be compatible with the SM prediction  after $1$ year of data taking with a $4$ kg detector (thus $1$ year $\times$ $4$ kg exposure). This allows us to compute sensitivity of CONUS on the production of $\chi$. More explicitly, we
adopt the following $\chi^{2}$-function comparing $N_{i}$ with $N_{i}^{{\rm SM}}$:
\begin{equation}
\chi^{2}=\sum_{i}\frac{[(1+a)N_{i}-N_{i}^{{\rm SM}}]^{2}}{\sigma_{{\rm stat},i}^{2}+\sigma_{{\rm sys},i}^{2}}+\frac{a^{2}}{\sigma_{a}^{2}}\,,\label{eq:cohdm-34}
\end{equation}
\begin{equation}
\sigma_{{\rm stat},i}=\sqrt{N_{i}+N_{{\rm bkg},\thinspace i}},\thinspace\sigma_{{\rm sys},i}=\sigma_{f}(N_{i}+N_{{\rm bkg},\thinspace i})\,.\label{eq:cohdm-35}
\end{equation}
Here $1+a$ is a rescaling factor with an uncertainty $\sigma_{a}=2\%$
which mainly comes from the overall uncertainty of the neutrino flux.
In addition, other systematic uncertainties  may change the shape
of the event spectrum, which is parametrized by $\sigma_{f}$ and
assumed to be $1\%$. The flux uncertainties used here are somewhat
optimistic. According to the previous theoretical calculations
\cite{Huber:2011wv,Mueller:2011nm,Kopeikin:2012zz}, the flux
uncertainty at 5 MeV is about 3\%. 
In the next few years, both the theoretical understanding and
experimental measurements will be considerably improved \cite{Buck:2015clx,Giunti:2016elf,Huber:2016xis} so
that the flux will be determined more precisely. 
The background $N_{{\rm bkg},\thinspace i}$
in each bin is 1 ${\rm count}/({\rm day}\cdot{\rm keV}\cdot{\rm kg})$.
For the nucleus recoil threshold we take 1.2 keV. 

The results are presented in Fig.~\ref{fig:constraints} where we show the
constraints in the $\bar{y}-m_{S}$ plane (with $m_{\chi}$ fixed)
and the $\bar{y}-m_{\chi}$ plane (with $m_{S}$ fixed). In the
$\bar{y}-m_{S}$ panels, the bounds are almost flat when $m_{S}<2$
MeV (CONUS) or $m_{S}<10$ MeV (COHERENT), which can be understood
from Eq.~(\ref{eq:cohdm-20}) where, for small $m_{S}$, $2MT$ dominates
over $m_{S}^{2}$ in the denominator. Similarly, in the $\bar{y}-m_{\chi}$ plots, 
the bounds are also approximately flat for small $m_{\chi}$ which
can be understood from the $K$ factor in Eq.~(\ref{eq:cohdm-21}). 
  However, the large mass behaviors are different for $m_{\chi}$
and $m_{S}$. As shown in the left panels of Fig.~\ref{fig:constraints},
the curves are approximatively linear for large $m_{S}$ because in this case the cross section is proportional to $(\bar{y}/m_{S})^{4}$. 
On the other hand, large $m_{\chi}$ can only be constrained by the
events with high $E_{\nu}$. If $m_{\chi}$ is larger than the maximal
value of $E_{\nu}$ of the neutrino flux, then there will be no constraint
at all because  neutrinos do not have sufficient energy to produce
$\chi$. For reactor neutrinos, the event rate above $6$ MeV is essentially too low to
have a significant impact and hence the sensitivity to the new physics scenario diminishes. Therefore, the 
CONUS
curves in the right panel rise up quickly around 6 MeV. For COHERENT,
the maximal $E_{\nu}$ is about $53$ MeV (half of $m_{\mu}$) but,
unlike in CONUS, the flux is not suppressed when $E_{\nu}$ is approaching
$53$ MeV, so the curves do not rise so quickly  when $m_{\chi}$
is close to the maximal $E_{\nu}$.

\begin{figure}[t]
	\centering
\includegraphics[width=7.5cm]{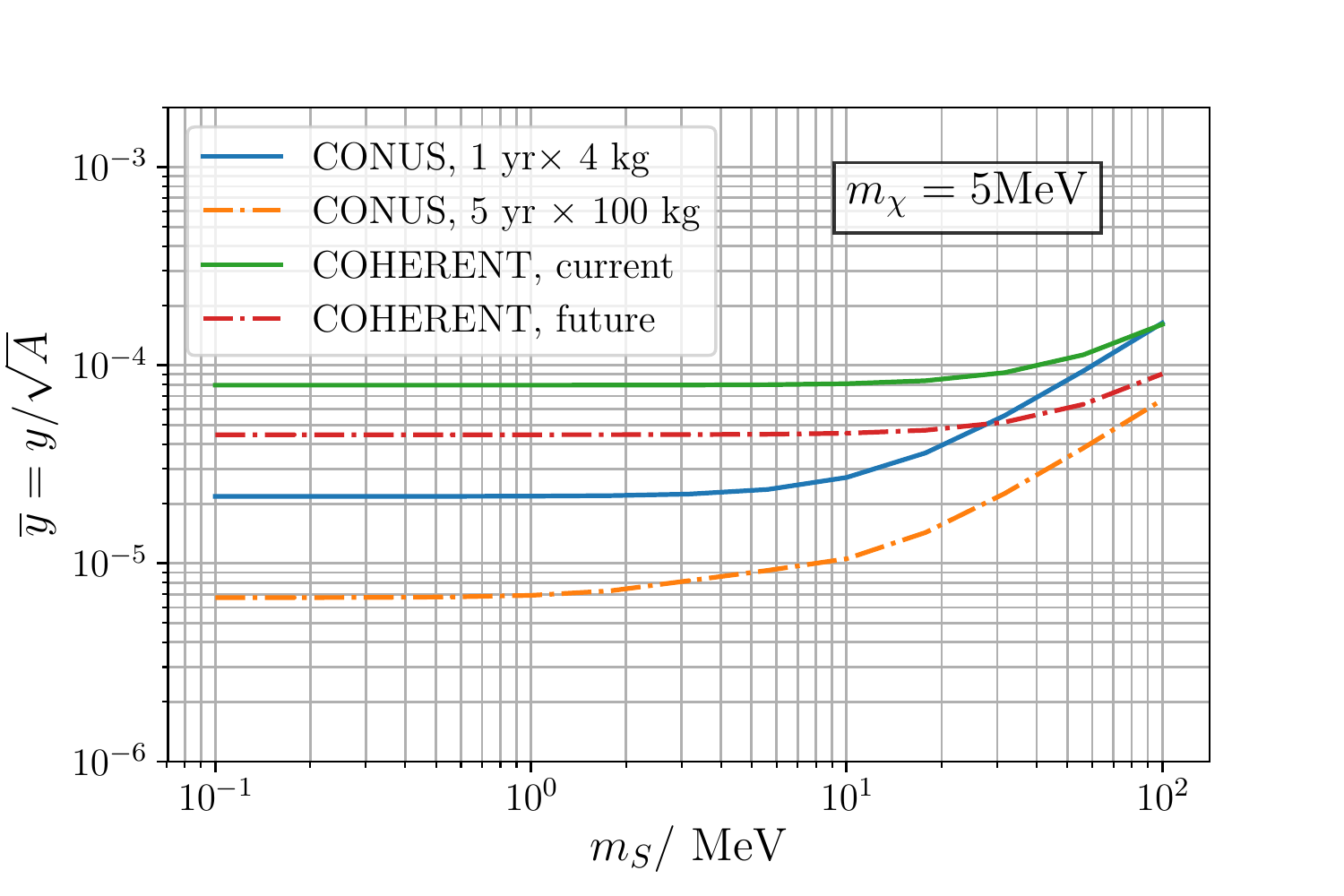}
	\caption{Future sensitivities of CONUS (5 yr $\times$ 100 kg exposure, 0.1 keV threshold) and COHERENT (statistics $\times$ 100)  on the benchmark $m_{\chi}=5$ MeV. For more details about the future configurations, see the text.}
\label{fig:future}
\end{figure}

In the future, the measurement of CE$\nu$NS will be significantly
improved by lower thresholds, larger fiducial masses, and longer exposure
times, etc. For reactor neutrinos, lower thresholds can increase the
statistics drastically because the current threshold actually only allows CONUS to 
measure the high energy tail of the reactor neutrino flux. For
COHERENT, using lower threshold detectors will not improve the measurement
significantly. This is because, unlike reactor neutrinos, the majority
of the neutrinos produced by $\mu^+$ or $\pi^+$ decays are not
in the low-energy region---see Fig.~\ref{fig:signal} for comparison. Consequently, lower thresholds for COHERENT cannot enhance the event numbers considerably and thus cannot improve the sensitivity significantly.
We will consider here the following two benchmark configurations 
to illustrate the future sensitivities of CE$\nu N$S experiments. 
The first one is running CONUS for 5 years with 100 kg Ge, and a considerably
improved threshold down to $0.1$ keV. In addition, the theoretical
uncertainties of reactor neutrino flux are assumed to be reduced by
a factor of 2. The second is (instead of doing a very detailled study
of various other detectors and target materials that are planed 
\cite{COH-talk}) 
increasing the statistics of COHERENT
by a factor of 100, which could be achieved by, e.g., a 20 times larger
fiducial mass with 5 times longer exposure. The systematic uncertainties
are correspondingly reduced so that we assume the overall uncertainty
is reduced by a factor of $\sqrt{100}=10$. In Fig.~\ref{fig:future},
we show the sensitivities of these two future experiments together
with their current constraints/sensitivities. Here, $m_{\chi}$ is
set at 5 MeV as a benchmark value. 

Let us now discuss other limits on the scenario under study. 
Regarding CE$\nu N$S, aspects of light scalars coupling to neutrinos and nuclei were explored in Ref.\  
\cite{Farzan:2018gtr}. Since in our framework a massive MeV-scale fermion $\chi$ is involved, most  
limits are expected to be 
weaker than the ones collected in Ref.\ \cite{Farzan:2018gtr}, where only couplings to nuclei and 
light neutrinos were considered. It is in addition 
 more complicated to obtain precise limits, so we focus here on 
giving reasonably robust estimates. 
It was found in Ref.\ \cite{Farzan:2018gtr} that 
all limits from terrestrial experiments, e.g.\ 
$n$-Pb scattering and meson decay experiments, are weaker than the bound from COHERENT 
as well as the CONUS sensitivity. 
BBN constraints, however, are relevant for $\mathcal{O}(1)$ MeV-scale $S$, and thus 
the $m_S=0$ curves in \cref{fig:constraints} should be interpreted as an illustration 
to show the strength of the limit in the small mass regime. 
When considering the $\chi$ density evolution in the early Universe (see \cref{sec:DM}), 
we will actually take $m_S \approx 10-100$ MeV. 

We should also mention limits from Supernova 1987A. If efficiently produced, the light states 
can carry a significant amount of energy from the Supernova core. In such case, the amount of 
energy carried by active neutrinos would be too small to match the observation of 
Supernova 1987A and hence a limit can be set. The leading process for the production of 
$S$ is $\nu\nu \to SS$ via $t$-channel $\chi$ exchange and it is suppressed by the fourth power of the 
small coupling $y_\chi$. 
As $\chi$ is concerned, in Ref.\ \cite{Chang:2018rso} the authors presented, within a specific model, 
that the cross section for scattering of a new light fermion on protons and electrons is constrained by 
Supernova 1987A cooling arguments to values comparable to the corresponding cross sections for neutrinos. 
This can be understood as follows: if a novel fermion acts as a fourth neutrino species inside of the star, 
it will carry away energy comparable to the one carried away by the individual active neutrino species. 
This suggests that $\sim 25$\% of the energy budget would be carried away by $\chi$. 
Given the astrophysical uncertainties associated to Supernova 1987A, 
exotic particles can carry away up to 50\% of the total energy of 
the collapse \cite{Davidson:2000hf,Foot:2015sia}. This corresponds again roughly to a new cross section 
of similar magnitude as the SM one. The reachable parameter values from \cref{fig:constraints} fulfill this 
constraint.  Hence, we infer that the cooling arguments are not 
excluding the relevant parameter space. Finally, note that $\chi$, being an MeV-scale particle, can 
not be resonantly produced through an MSW effect \cite{Wolfenstein,Mikheev:1986gs,Mikheev:1986wj}. 
Such effect is very relevant for keV-scale particles for 
which strong limits can be derived \cite{Arguelles:2016uwb}. 
\section{\boldmath{$\chi$} and Neutrino Mass Generation}
\label{sec:model}

\noindent
In this section we will discuss the possible connection of the new fermion $\chi$ to 
neutrino mass generation. Any fermion that couples to light active
neutrinos must be investigated with regard to its contribution to
neutrino mass. \\


Let us first discuss the nature of the scalar $S$ that appears in our
framework. 
Given our preference for light $S$, such construction is 
not achievable with representations higher than singlets. 
Namely, $S$ can obviously not be the SM Higgs due to its tiny couplings with $u$ and $d$ quark as well as its heavy mass which would further suppress the strength of the 
CE$\nu N$S process. If we replace the SM Higgs by a novel Higgs doublet $\Phi$ 
with possibly larger couplings to quarks (and hence nuclei), we face
the problem of a neccessary huge mass splitting between the light 
neutral component and the charged ones, which have not been seen. 
An option would be to consider the following gauge invariant
$5$-dimensional operator in the 
effective theory formalism
\begin{align}
\mathcal{L} \supset \frac{1}{\Lambda}\, \left(S \bar{\chi}\right) \left(\tilde{H}^\dagger L\right) + \text{h.c.},
\label{eq:5D}
\end{align}  
with singlets $S$ and $\chi$, 
where $\Lambda$ represents the scale of new physics. After electroweak symmetry breaking
 this operator yields an interaction term 
\begin{align}
\frac{v}{\Lambda} \,S \bar{\chi} \,\nu\,.
\label{eq:eff_coupling}
\end{align} 
By assuming furthermore non-vanishing interactions between $S$ and
nuclei (or light quarks), the CE$\nu N$S occurs through
the process shown in \cref{fig1}. 
We will now discuss a minimal model containing SM singlets only, which will
generate the effective Lagrangian in Eq.\ (\ref{eq:5D}).

\subsection{The Model}
\label{subsec:model}
\noindent
We supplement the SM particle content with
\begin{align}
&\chi\sim (1,1,0)\,, &  N_R  &\sim (1,1,0)\,,  &  S &\sim (1,1,0)\,,
\end{align}
where $\chi$ and $N_R$ are Majorana fermions and $S$ is a real
scalar. The quantum numbers under the SM gauge group
$SU(3)_c\times SU(2)_L \times U(1)_Y$ are indicated in brackets, and
clearly no charged degrees of freedom are introduced. One of the goals
of this section is to demonstrate that the neutrino masses can be
generated from this extended fermion sector via a modified type-I
seesaw mechanism
\cite{Goran,Minkowski,GellMann:1980vs,Yanagida:1979as}. This means
that we would require at least two generations of novel fermions which
participate in this mechanism, such that at most one  light neutrino
is massless. Still, for simplicity, throughout this section we will
focus on the $1$-generation case which can be straightforwardly
extended. Similarly, we will also restrict our discussion to one
active neutrino flavor state, 
namely for definitness the electron (anti)neutrino $\parenbar{\nu_e}$. 

\begin{figure*}[t!]
	\centering
\includegraphics[width=0.35\textwidth]{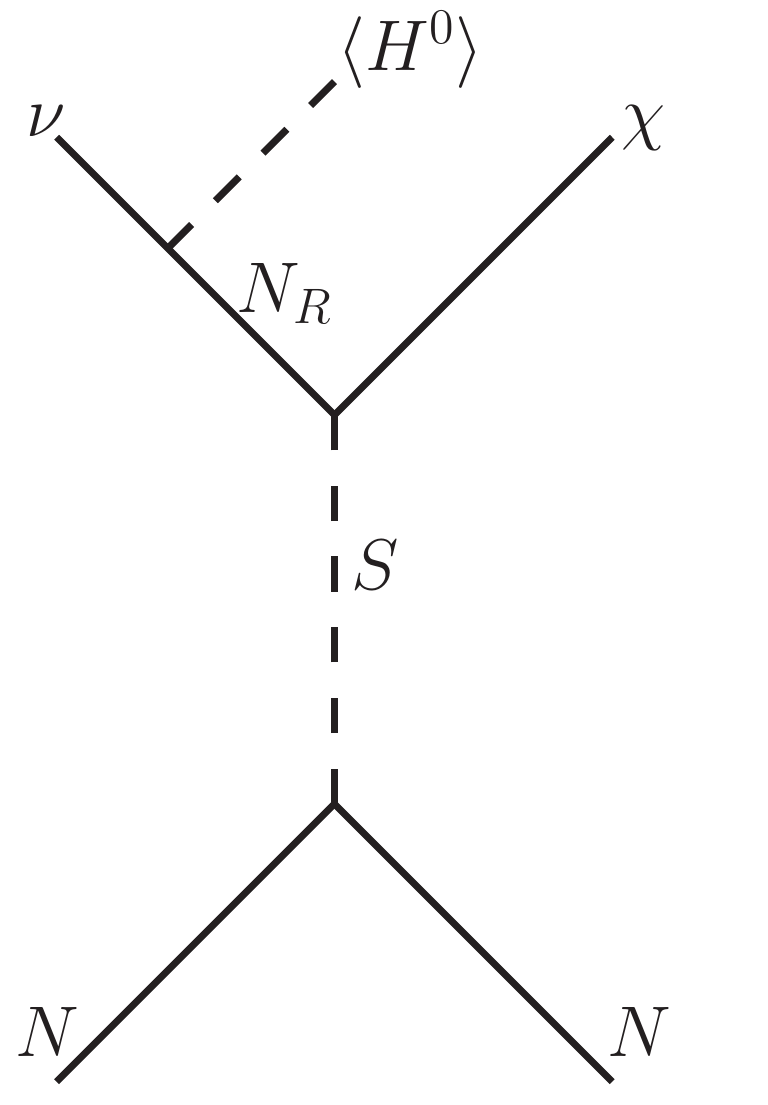}
	\caption{The process for CE$\nu N$S in the UV complete realization given in \cref{subsec:model}.}
\label{fig2}
\end{figure*}

The relevant part of the Lagrangian reads
\begin{align}
\mathcal{L}\supset y_1 \bar{N}_R \tilde{H}^\dagger L + \frac{1}{2} M_N \bar{N}_R N_R^c +
y_2 \,\bar{\chi} \tilde{H}^\dagger L + \frac{1}{2} m_\chi \bar{\chi}\, \chi^c +
y_3 \,\bar{\chi} S N_R^c + M_1 \bar{N}_R \,\chi^c + \text{h.c.},
\label{eq:lagr}
\end{align}
where $y_i~(i=1,2,3)$ are  the Yukawa couplings\footnote{In this work we do not study CP violation in the lepton sector so these couplings can be taken real for simplicity.}, $m_\chi$ and $M_N$ are Majorana masses of $\chi$ and $N_R$ fields, respectively, and $M_1$ is the Dirac mass which is allowed by gauge symmetries. \\
This Lagrangian is a minimal UV complete realization of \cref{eq:5D,eq:eff_coupling} with fermion singlet $N_R$ interacting with the fields given in both brackets of \cref{eq:5D} through Yukawa couplings $y_1$ and $y_3$ (see \cref{eq:lagr}). We will show that the allowed values of $M_N$ exceed the characteristic momentum exchange $q^2$ in CE$\nu N$S experiments, which justifies the analysis setup in \cref{sec:production}. 
If $M_N^2\gg q^2$, we can easily relate the parameters of the full
theory with $\Lambda$ and obtain $\Lambda = M_N/(y_1 y_3)$. If that
was not the case, the topology shown in \cref{fig2} 
including $N_R$ as the dynamical degree of freedom should be considered. 

More importantly, within the presented model, we will demonstrate the
existence of parameter space that can be probed by CE$\nu N$S
experiments, generates neutrino masses in the right ballpark, and is not excluded from the new physics searches at neutrino oscillation facilities, beam dump experiments, colliders, etc. This indicates the importance of the CE$\nu N$S in future new physics searches as there are scenarios where it could yield the strongest limits or perhaps even lead to new discoveries.

After electroweak symmetry breaking, the neutral fermion mass matrix reads
\begin{align}
\begin{pmatrix} \bar{\nu}_L^c & \bar{\chi} & \bar{N}_R \end{pmatrix}
\begin{pmatrix}
0  & y_2 v  & y_1 v  \\[0.05cm]
 y_2 v  & m_\chi  & M_1 \\[0.05cm]
y_1 v  & M_1 & M_N 
\end{pmatrix}
\begin{pmatrix} \nu_L \\ \chi^c \\ N_R^c \end{pmatrix},
\label{eq:neutral_mass}
\end{align} 
where $v \equiv\langle H^0 \rangle=174$ GeV and we assumed that $S$ does not develop a non-vanishing vacuum expectation value.

We furthermore assume for the mass matrix given in \cref{eq:neutral_mass} that
$M_1\ll  M_N$. In this way, the mixing between $\chi$ and $N_R$ is
suppressed and hence, the masses of heavy new fermions essentially match the parameters in the flavor basis, $m_\chi$ and $M_N$. Contrary, if $M_1 \simeq M_N$, the two physical masses would be of similar size which is not wanted in our scenario.

We start by performing a rotation in the $1$-$2$ plane by an angle
$\theta_{e\chi}= y_2 \,v /m_\chi\ll 1$. As discussed above, $m_\chi$
is the physical mass of a particle produced in CE$\nu N$S
experiments. We take  $m_\chi=5$ MeV as an illustrative number. The
bounds on the mixing of active neutrinos with heavy fermions have been
extensively studies in the literature. From Refs.\ \cite{Atre:2009rg,deGouvea:2015euy,Deppisch:2015qwa}
we infer that the constraint on the mixing between $\nu_e$ and $\chi$ for $m_\chi=5$ MeV reads\footnote{Somewhat stronger cosmological limits exist in the literature \cite{Vincent:2014rja}. Unlike laboratory limits, these can be evaded by assuming for instance a novel decay channel for $\chi$.}
\begin{align}
\theta_{e\chi}^2 & \leq 10^{-7} \hspace{5mm}\Longrightarrow \hspace{5mm}y_2 \lesssim 9\times 10^{-9}\,,
\label{eq:limit}
\end{align}
and is set by neutrinoless double beta decay experiments.
Weaker
limits apply for the other flavors, which therefore can be accommodated
more easily. 
The mass matrix after the $1$-$2$ rotation reads approximately
\begin{align}
\mathcal{M}=\begin{pmatrix}
-y_2^2 \,v^2 /m_\chi  & 0  & y_1 \,v  \\[0.05cm]
 0  & m_\chi  & y_1 y_2 \, v^2 / m_\chi \\[0.05cm]
y_1\, v  & y_1 y_2\, v^2 / m_\chi & M_N 
\end{pmatrix},
\label{eq:neutral_mass2}
\end{align} 
from where one can infer that $\chi$ may serve as a potential source
of neutrino mass. 
By taking the upper value of $y_2$ in \cref{eq:limit} we obtain $m_\nu
\approx y_2^2 \,v^2 /m_\chi \sim 0.1$ eV which matches the required
order of magnitude for neutrino mass.

It was demonstrated in \cref{sec:production} that the numerical
analysis of CE$\nu N$S 
for  $m_S \lesssim 100$ MeV yields the limit
\begin{align}
\bar{y}\equiv \sqrt{\frac{y_N}{A}\,y_\chi} = \sqrt{\frac{y_N}{A}\frac{v}{M_N}\, y_1 y_3} \lesssim \big[10^{-5}, 10^{-4}\big],
\label{eq:Y}
\end{align}
where $y_\chi$ was introduced in \cref{eq:eft-lagrangian} and $y_N/A$
roughly corresponds to the coupling strength to individual quarks. The
values indicated in square brackets represent the range in which the
bound, depending on specific values of $m_S$ and $m_\chi$,  is set
(see \cref{fig:constraints}). We assume $y_N/A\simeq y_\chi$, i.e.\
similar size of the 
$S$ coupling to quarks and fermions, such that 
\begin{align}
y_\chi \equiv \frac{v}{M_N}\, y_1 y_3 \lesssim \big[10^{-5}, 10^{-4}\big]
\label{eq:Y2}
\end{align}
approximatively holds. 
Having now a feeling for the numbers in \cref{eq:neutral_mass2} , we
continue the diagonalization. Performing a rotation in the $1$-$3$ plane
by an angle 
$\theta_{eN}= y_1 \,v /M_N\ll 1$ and using this expression as well as \cref{eq:Y2} we can relate the mixing angle with the upper limit from CE$\nu N$S experiments
\begin{align}
\theta_{eN} \simeq \frac{\big[10^{-5}, 10^{-4}\big]}{y_3}\,.
\label{eq:mix_eN}
\end{align}
Clearly, $y_3$ must not be tiny as otherwise the large mixing would pose a problem for $M_N \lesssim \mathcal{O}(\text{TeV})$. We can safely assume $y_3=\mathcal{O}(1)$ because it parametrizes the strength of the interaction between three hidden particles and moreover $S$ does not mix with the SM Higgs. By inserting $v=174$ GeV in \cref{eq:Y2} we obtain the relation 
\begin{align}
M_N \simeq y_1 \big[10^6, 10^7\big]\, \text{GeV}\,.
\label{eq:y1-M1}
\end{align}  
The seesaw contribution to the neutrino mass from mixing between $\nu_e$ and $N_R$ is
then 
\begin{align}
m_\nu \sim \frac{y_1^2 v^2}{M_N} \simeq y_1 \,\big[10^{-3}, 10^{-2}\big] \,\text{GeV}\,,
\label{eq:nu-mass-N}
\end{align}
which gives the upper bound on $y_1$ from neutrino mass considerations
\begin{align}
y_1 \lesssim \big[10^{-8}, 10^{-7}\big]\,.
\label{eq:y_1}
\end{align}
From \cref{eq:y1-M1,eq:y_1} we infer that the $M_N$ values which can
contribute to this neutrino mass generation  are in the $\mathcal{O}(10^{-2}-1)$ GeV mass range. Finally, we need to check if the mixing given in \cref{eq:mix_eN} is compatible with such masses. To this end, we again employ the limits from 
Refs.\ \cite{Atre:2009rg,deGouvea:2015euy,Deppisch:2015qwa} and infer
that $M_N = \mathcal{O}(1)$ GeV is fully consistent, whereas the
smaller values are marginally allowed, i.e.\ in tension with the
constraints from neutrinoless double beta decay experiments, big bang
nucleosynthesis as well as the PS191 \cite{Bernardi:1985ny} beam dump
experiment. Interestingly, GeV-scale $M_N$ will be testable at some upcoming experiments such as  DUNE
\cite{Acciarri:2016crz}, SHiP \cite{Lantwin:2017xtc}, FASER \cite{Feng:2017uoz, Feng:2017vli, Kling:2018wct}, NA62 \cite{CortinaGil:2017mqf} and MATHUSLA \cite{Curtin:2018mvb}. 

In summary, parameters that are compatible with all available laboratory constraints and give an
  observable signal in coherent scattering experiments, give a
  neutrino mass of order 
\begin{equation} \label{eq:mnu}
\left(\frac{m_\nu}{0.1\, \rm eV} \right) 
\approx (1-x)\left(\frac{y_1}{10^{-7.25}} \right)^2 \left(\frac{\rm GeV}{M_N} \right) + x \,
  \left(\frac{y_2}{10^{-8.75}} \right)^2 \left(\frac{\rm MeV}{m_\chi} \right),
\end{equation}
which is compatible with observation. Here, $x\in \left[0,1\right]$ denotes relative contribution to the active neutrino mass from $\chi$ and $N_R$.



\section{\boldmath{$\chi$} as Dark Matter Particle}
\label{sec:DM}
 \noindent
Limits on $m_S$ from terrestrial experiments as well as astrophysics
were discussed in \cref{sec:production}. This section is devoted to
the evaluation of the cosmic abundance of $\chi$. As a first
observation, we note that the smallness of $m_\chi$ relative to the
electroweak scale implies that the production of DM from freeze-out
might not yield desired results and signifies the preference for
non-thermal production. Note further that within our framework there is no possibility for $\chi$ to decay into a pair of electrons or neutrinos. However, there could be tree-level ($\chi\to 3\nu$) and radiative decays ($\chi \to \nu \gamma$). From the expressions for the decay rates of these processes
\cite{Xing:2011,Arguelles:2016uwb} we infer that in order to ensure
$\chi$ stability, $\theta_{e\chi} \lesssim 4.8\cdot 10^{-9}$ for
$m_\chi=5$ MeV. The implied tiny value of $y_2$ does not jeopardize our new
physics scenario at CE$\nu N$S  experiments because the rate for $\nu
N \to \chi N$ depends on $y_1$ and $y_3$ couplings, but not on $y_2$. 
Note however from Eq.\ (\ref{eq:mnu}) that neutrino mass will be dominated by $M_N$ in this case. \\

If $\chi$ is the DM particle, CE$\nu N$S experiments would yield an entirely novel method for searching (MeV-scale) DM. 
Note that dark matter in CE$\nu N$S experiments was discussed  
already in \cite{deNiverville:2015mwa,Ge:2017mcq}, where kinetically mixed dark photons decay into 
DM pairs which subsequently scatters at CE$\nu N$S experiments (note that the neutrino sources 
of those experiments also generate photons).  
Here, we will propose a novel framework by demonstrating that the 
DM particle can be produced directly via CE$\nu N$S and the effects of such process are imprinted in the measurable recoils of the nuclei. 
Interestingly, this would resemble DM search in direct detection experiments because in both cases the observable signal is the nuclear recoil. While nuclear recoil in direct detection experiments stems from DM-nucleus interaction, in CE$\nu N$S such effect is caused by neutrinos. Thus, the main goal would be to distinguish between the SM coherent neutrino scattering events and ``new physics" events in which $\chi$ is produced. We have shown in \cref{sec:production} that the distributions of the recoil energy corresponding to these two cases can be vastly different and it is possible to discriminate between them. These arguments strongly motivate the search for DM at CE$\nu N$S experiments if $\chi$ can indeed play the role of DM. 
Hence, in what follows we will discuss if and how $\chi$ can be produced in right amounts. 
We rely on the UV complete model introduced in \cref{sec:model} which was already shown to be successful in generating non-vanishing neutrino masses.

\begin{figure}[t!]
	\centering
	\includegraphics[width=0.48\textwidth]{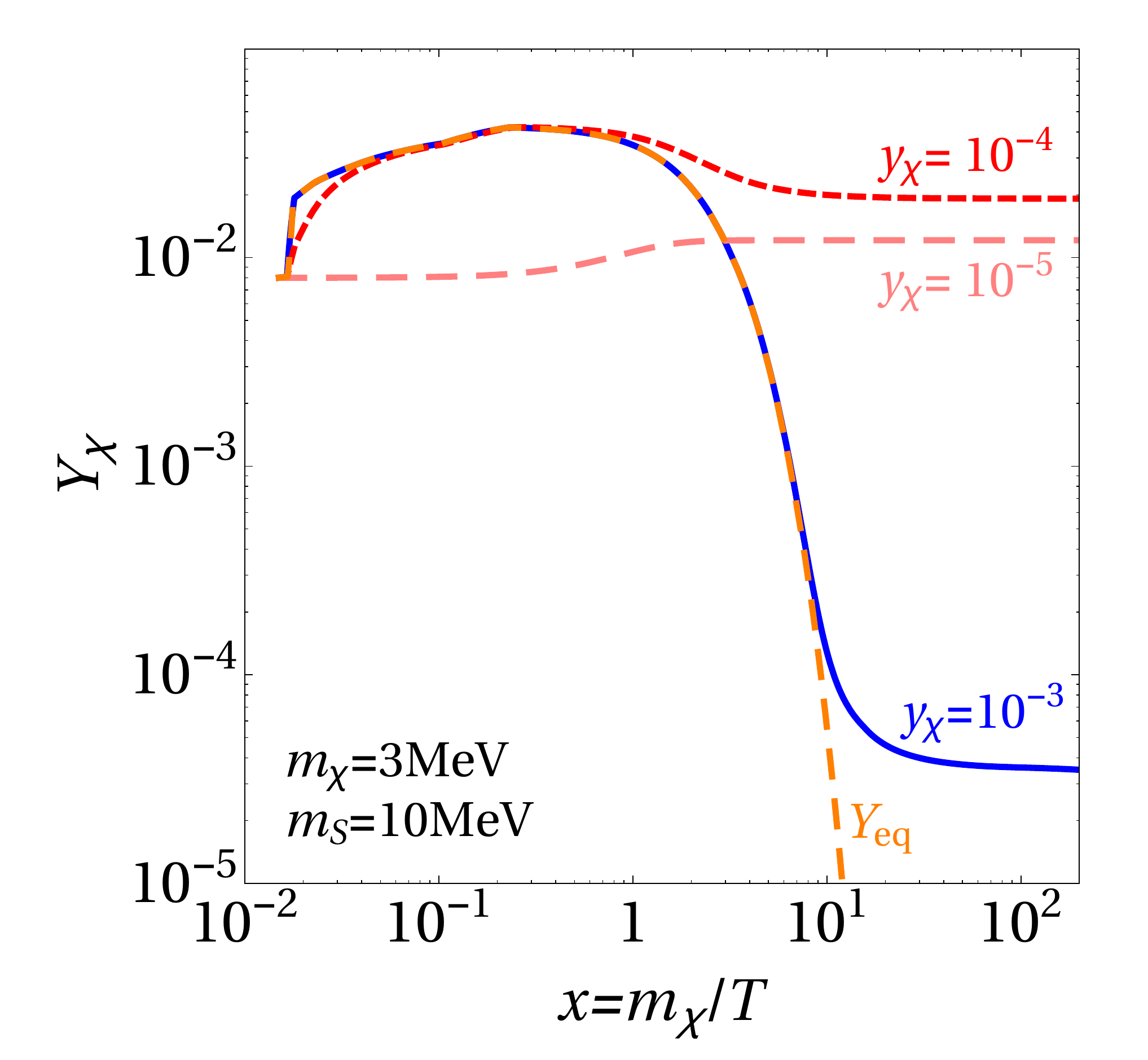}
	\includegraphics[width=0.48\textwidth]{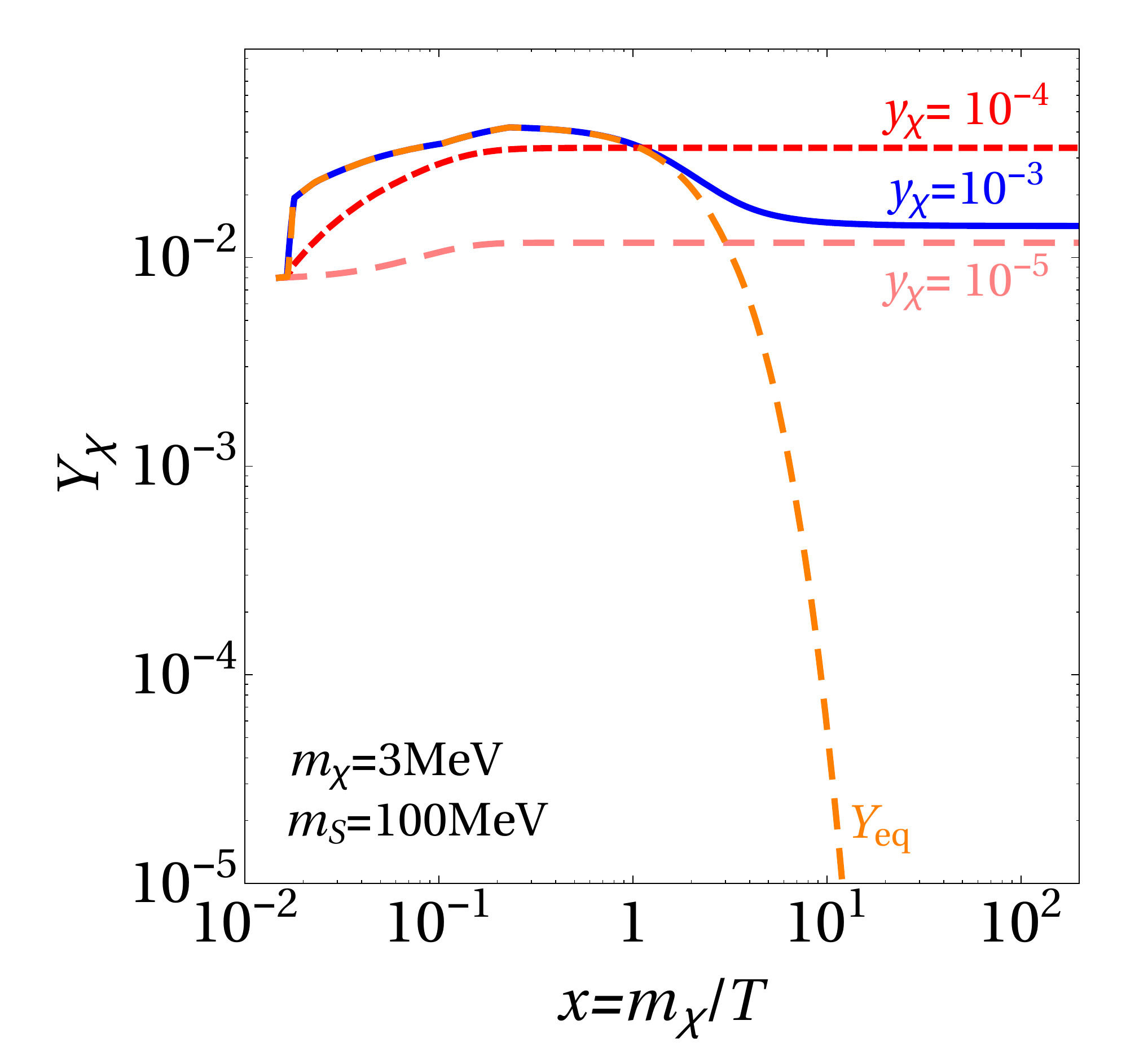}
	\caption{In left (right) panel we show the evolution of $Y_\chi$ for several different values of $y_\chi$ with fixed $m_\chi=3$ MeV and $m_S=10$ MeV $(m_S=100\,\text{MeV})$.  For larger couplings, characteristic freeze-out curves are easy to identify whereas $\chi$ is ``frozen-in" for weaker $\chi\chi \leftrightarrow \bar{\nu}\nu$ interaction. In the latter scenario, the $\chi$ abundance is set primarily by the initial condition, i.e.\ not-vanishing $Y_\chi$ at $T_\text{QCD}$.}
	\label{fig:DM}
\end{figure}

In order to realize the CE$\nu N$S process shown in \cref{fig1} we require $S$ to interact with quarks as well as  with $\chi$ and neutrinos. The Lagrangian 
for these interactions reads 
\begin{align}
\mathcal{L} \supset y_{sqq} S\bar{q}q + y_\chi S \bar{\chi} \nu \,,
\label{eq:lag_Yukawa}
\end{align}
where  $y_{sqq}\simeq y_N/A$ and $y_\chi$  are the couplings to quarks
and leptons, respectively. We consider the case in which neither of
these couplings is weaker than $\mathcal{O}(10^{-6})$. 
For an UV complete model from which \cref{eq:lag_Yukawa} stems after
electroweak symmetry breaking, we refer the reader to \cref{sec:model,sec:production}.

 For $m_S \lesssim 100$ MeV and $y_{sqq}\ge \mathcal{O}(10^{-6})$, the 
 decay and inverse decay widths of the process $S\leftrightarrow 
 \bar{q}q$ are much larger than the value of the Hubble parameter  
 across the relevant temperatures. Thus, $S$ is in thermal equilibrium
 with the SM bath. At the temperatures below the QCD phase transition
 ($T_\text{QCD}\simeq 200$ MeV) quarks are no longer relevant degrees
 of freedom. The vast majority of the formed mesons and baryons are
 non-relativistic and hence effectively disappear from the SM thermal
 bath shortly below $T_\text{QCD}$ (the only exception are pions which
 are somewhat lighter). The $S\leftrightarrow \bar{q}q$ interaction
 which was keeping $S$ in  thermal equilibrium is not relevant below
 $T_\text{QCD}$. However, even below $T_\text{QCD}$, $S$ can be in
 thermal equilibrium through the process $SS\leftrightarrow
 \bar{\nu}\nu$, provided there is a sufficiently large Yukawa coupling
 $y_\chi$. Such process occurs via $t$-channel exchange of $\chi$ and
 is governed exclusively by the second term in
 \cref{eq:lag_Yukawa}. Note that $S$ should decay shortly below
 $T_\text{QCD}$ into $\chi$ and neutrinos in order not to violate any
 of the BBN predictions. For the parameter values we will take below this is indeed the case. 
 It is also worthwhile to mention that the $N_R$, introduced in
 \cref{sec:model} to UV-complete the model, decay well above
 $T_\text{QCD}$ due to their large $y_3$ coupling. On the other hand,
 the stability of $\chi$ is ensured by assuming a small value of $y_2$ which sufficiently suppresses $\chi\to 3\nu$ and $\chi\to \nu\gamma$ decays.

Above $T_\text{QCD}$, $\chi$ is in thermal equilibrium with $S$ and
hence also with the SM bath, due to rapid $S\leftrightarrow \chi \nu$
processes. Whether $\chi$ remains in contact with the SM bath at lower
temperatures (after $S$ decays) depends on the strength of the $\bar{\chi}\chi\leftrightarrow \bar{\nu}\nu$ process which occurs via $t$-channel exchange of $S$ (hence is proportional to $y_\chi^4$).\\

In order to accurately determine the present-day abundance of $\chi$  we solve the following Boltzmann equations \cite{Garny:2017rxs,Ellis:2015vaa}
\begin{align}
\frac{dY_\chi}{d x} =& \frac{1}{3 H} \frac{ds}{dx} \bigg[ \langle \sigma v \rangle_{\bar{\chi}\chi\to \bar{\nu}\nu} \left(Y_\chi^2-(Y_\chi^\text{eq})^2\right)
-\frac{\Gamma_S}{s} \frac{K_1(\frac{m_S}{m_\chi} x)}{K_2(\frac{m_S}{m_\chi} x)}
\left(Y_S -Y_\chi \frac{Y_S^\text{eq}}{Y_\chi^\text{eq}}\right)\nonumber \\ &
+\langle \sigma v \rangle_{\bar{\chi}\chi\to SS } \left(Y_\chi^2 -Y_S^2 \left(\frac{Y_\chi^\text{eq}}{Y_S^\text{eq}}\right)^2\right)
\bigg]\,,\nonumber \\ 
\frac{dY_S}{d x} =& \frac{1}{3 H} \frac{ds}{dx} \bigg[ \langle \sigma v \rangle_{SS\to \bar{\nu}\nu} \left(Y_S^2-(Y_S^\text{eq})^2\right)
+\frac{\Gamma_S}{s} \frac{K_1(\frac{m_S}{m_\chi} x)}{K_2(\frac{m_S}{m_\chi} x)}
\left(Y_S -Y_\chi \frac{Y_S^\text{eq}}{Y_\chi^\text{eq}}\right)\nonumber \\ &
-\langle \sigma v \rangle_{\bar{\chi}\chi\to SS } \left(Y_\chi^2 -Y_S^2 \left(\frac{Y_\chi^\text{eq}}{Y_S^\text{eq}}\right)^2\right)
\bigg]\,.
\label{eq:Boltzmann}
\end{align}  
For a particle denoted with $i$ we define the yield as $Y_{i}=n_{i}/s$
(superscript eq denotes equilibrium value), where $n_i$ and $s$
are number and entropy density, respectively. In addition, $x=m_{\chi}/T$, $H$ is the Hubble parameter, $\langle \sigma v \rangle$ is the thermally averaged cross section for a given process (evaluated following Ref.\ \cite{Gondolo:1990dk}), $\Gamma_S=y_\chi^2 m_S/(8\pi)$, $K_1$ and $K_2$ are modified Bessel functions. The initial conditions are $Y_\chi(T_\text{QCD})=Y_\chi^\text{eq}$ and 
$Y_S(T_\text{QCD})=Y_S^\text{eq}$, since we argued above that the sizable coupling to free quarks 
leaves the particles in thermal equilibrium at temperatures above $T_\text{QCD}$. 

The solution for $Y_\chi(x)$ is shown in \cref{fig:DM}. In both panels we set
$m_\chi= 3$ MeV and show the results for $y_\chi=\{10^{-3},\,10^{-4},\,10^{-5}\}$ which are in the ballpark of testable values at CE$\nu N$S experiments. In the (left) right panel, $m_S$ is fixed to $10$ MeV ($100$ MeV). Much smaller values of $m_S$ are not considered due to the requirement $m_\chi<m_S$ which forbids $\chi$ to decay and is thus essential to render our DM candidate  stable. We discuss and interpret the results from the figure in what follows. 

In both panels we observe the standard thermal freeze-out of $\chi$ for $y_\chi=10^{-3}$. 
The interaction which keeps $\chi$ in thermal equilibrium with the SM is $\bar{\chi}\chi\to \bar{\nu}\nu$, and the strength of this process increases by reducing $m_S$. Hence, for smaller $m_S$, the stronger interaction indicates that $\chi$ stays longer in the thermal equilibrium and undergoes freeze-out at later times (smaller temperatures) which implies smaller final abundance of $\chi$. This is visible from \cref{fig:DM} where the final value of $Y_\chi$ for $y_\chi=10^{-3}$ (blue line) in the left panel (lighter $S$) is much smaller that the one shown in the right panel. By using the relation between the yield and the relic abundance 
\begin{align}
\Omega h^2=2.742\cdot 10^5 \left(\frac{m_\chi}{1\,\text{MeV}}\right) Y_\chi\,,
\label{eq:rel}
\end{align} 
it is clear from both panels of \cref{fig:DM} that $\chi$ is strongly overproduced with respect to the observed DM abundance $\Omega h^2 \approx 0.12$ \cite{Aghanim:2018eyx}. 
Moreover, the freeze-out temperatures are smaller than $\mathcal{O}(1)$ MeV which means 
that any new physics contribution invoked to deplete $\chi$ is
strongly constrained by BBN considerations. 

The freeze-out also occurs
for $m_\chi=10$ MeV and $y_\chi=10^{-4}$ (red line in the left
panel). In this case, initially, the interaction between $\chi$ and
the SM is not sufficiently strong to follow the sudden change of
$Y_\chi^{\rm eq}$ which is induced by a rapid change of SM degrees of freedom present in the thermal bath at $T= \mathcal{O}(10-100)$ MeV. However, $Y_\chi$ eventually reaches the equilibrium value and the freeze-out occurs leaving 
$\chi$ with even larger abundance than in the previously discussed scenarios. For
 $m_\chi=100$ MeV and $y_\chi=10^{-4}$ (red line in the right panel)
 as well as for both cases with $y_\chi=10^{-5}$, $\chi$ does not
 reach equilibrium. This means that the production occurs via
 ``freeze-in" \cite{Hall:2009bx} where the weak interaction with the SM
 bath leads to a gradual accumulation of $\chi$ abundance. From
 \cref{fig:DM} it is obvious that the final $Y_\chi$ in all
 ``freeze-in" scenarios is chiefly set by the initial condition
 (non-vanishing $Y_\chi$ at QCD phase transition) and the freeze-in
 contribution yields only a subdominant effect. \\
Even though the DM
 abundance is still too large, in the ``freeze-in" scenarios where
 $\chi$ is not in thermal equilibrium at $\mathcal{O}(1)$ MeV
 temperatures, a late-time entropy injection episode below
 $T_\text{QCD}$ can reduce the $\chi$ abundance significantly. Such
 entropy injection can be achieved for instance via decays of heavy
 scalars into the states in the thermal bath (see Ref.\ \cite{Dutra:2018gmv} and references therein). It
 is clear that $Y_\chi$ needs to be diluted by
 $\mathcal{O}(10^4-10^5)$ in order to meet observation. 
Let us note that alternatives with respect to late-time entropy
injection have also been considered. For instance, a heavier,
overproduced particle could decay into lighter states which may be DM
\cite{Molinaro:2014lfa,Baumholzer:2018sfb}. This helps because, as may
be inferred from \cref{eq:rel}, the abundance of DM is proportional to
the mass of a DM particle. A detailed analysis of such a scenario is
beyond the scope of this project as it would require a significant
extension of the Boltzmann equations given in \cref{eq:Boltzmann}. In such extension, the late-time entropy injection would be avoided. We also note that there may be more models in which MeV-scale particle is produced in right amounts,
without a necessity for late-time entropy injection. \\
 
 In conclusion, for the couplings that can be probed in CONUS or
 COHERENT 
 $(y_\chi\sim \big[10^{-5}, 10^{-4}\big])$, we explored whether in our 
 model $\chi$ can be DM, i.e.\ produced in right amounts in the early Universe. It turns out (see again \cref{fig:DM}) that this is not achievable without extending the minimal model presented in \cref{sec:model}. Namely, we find that a late time entropy injection episode is necessary to sufficiently deplete the abundance of $\chi$ and render it a viable DM candidate.\\
Since $\chi$ in our model does not scatter on nuclei and electrons at tree-level, the bounds from direct detection are not strong. Moreover, for the considered range of couplings and masses, the CE$\nu N$S cross section is several orders of magnitude smaller than those that can be currently tested at direct DM detection facilities for  MeV-scale DM \cite{Essig:2015cda}. Given the absence of the annihilation channel into $e^+e^-$ pairs, the constraints from CMB \cite{Slatyer:2015jla} are also not probing the parameter space to which CE$\nu N$S experiments are sensitive. 

\section{Summary and Conclusions}
\label{sec:summary}
 \noindent
Coherent neutrino-nucleus scattering is a new window to probe physics
within and beyond the Standard Model. 
We have noted here that the final
state fermion does not necessarily have to be a light active
neutrino. Instead, we have entertained  the possibility that an MeV-scale
fermion $\chi$ is produced in the process, which will lead to a significant
modification of the observable recoil spectrum. We have set limits on
the parameters that are involved when the interaction of the 
neutrino-$\chi$ pair with quarks is mediated by a light scalar. 

The measurable couplings are well compatible with
neutrino mass generation via low-scale type-I seesaw mechanism where, interestingly, 
both $\chi$ and the newly introduced GeV-scale fermion $N_R$ can contribute. 
Furthermore, $\chi$ may be
the DM particle, which we have shown to be typically requiring an
injection of entropy in the early Universe after the QCD phase transition.
Such an entropy injection can be achieved by introducing a new scalar which 
decays to the states in the thermal bath at late times, diluting the dark matter to the abundance which is in accord 
with present observations.

Thus, exotic physics in coherent  neutrino-nucleus scattering has a
variety of interesting implications in neutrino physics and
cosmology. The present analysis is only one example of the exciting
prospects that this new window to physics has given us the opportunity
to probe. 

\section*{Acknowledgments}
\noindent
We would like to thank Giorgio Arcadi, Yasaman Farzan, Rasmus S.L.\ Hansen and Stefan Vogl for useful discussions. 
WR is supported by the DFG with grant RO 2516/7-1 in the Heisenberg program.  
\bibliographystyle{JHEP}
\bibliography{refs}

\providecommand{\href}[2]{#2}\begingroup\raggedright\begin{thebibliography}{10}

\bibitem{Akimov:2017ade}
{\bf COHERENT} {\bf Collaboration}, D.~Akimov {\em et~al.}, {\it {Observation
  of Coherent Elastic Neutrino-Nucleus Scattering}},  {\em Science} {\bf 357}
  (2017), no.~6356 1123--1126, [\href{http://www.arxiv.org/abs/1708.01294}{{\tt
  1708.01294}}].

\bibitem{Freedman:1973yd}
D.~Z. Freedman, {\it {Coherent Neutrino Nucleus Scattering as a Probe of the
  Weak Neutral Current}},  {\em Phys. Rev.} {\bf D9} (1974) 1389--1392.

\bibitem{Horowitz:2003cz}
C.~J. Horowitz, K.~J. Coakley, and D.~N. McKinsey, {\it {Supernova observation
  via neutrino - nucleus elastic scattering in the CLEAN detector}},  {\em
  Phys. Rev.} {\bf D68} (2003) 023005,
  [\href{http://www.arxiv.org/abs/astro-ph/0302071}{{\tt astro-ph/0302071}}].

\bibitem{Drukier}
A.~Drukier and L.~Stodolsky, {\it Principles and applications of a
  neutral-current detector for neutrino physics and astronomy},  {\em Phys.
  Rev. D} {\bf 30} (Dec, 1984) 2295--2309.

\bibitem{Anderson-nus}
A.~J. Anderson, J.~M. Conrad, E.~Figueroa-Feliciano, C.~Ignarra, G.~Karagiorgi,
  K.~Scholberg, M.~H. Shaevitz, and J.~Spitz, {\it Measuring active-to-sterile
  neutrino oscillations with neutral current coherent neutrino-nucleus
  scattering},  {\em Phys. Rev. D} {\bf 86} (Jul, 2012) 013004.

\bibitem{Dutta-nus}
B.~Dutta, Y.~Gao, A.~Kubik, R.~Mahapatra, N.~Mirabolfathi, L.~E. Strigari, and
  J.~W. Walker, {\it Sensitivity to oscillation with a sterile fourth
  generation neutrino from ultralow threshold neutrino-nucleus coherent
  scattering},  {\em Phys. Rev. D} {\bf 94} (Nov, 2016) 093002.

\bibitem{Kosmos-nus}
T.~S. Kosmas, D.~K. Papoulias, M.~Tortola, and J.~W.~F. Valle, {\it {Probing
  light sterile neutrino signatures at reactor and Spallation Neutron Source
  neutrino experiments}},  {\em Phys. Rev.} {\bf D96} (2017), no.~6 063013,
  [\href{http://www.arxiv.org/abs/1703.00054}{{\tt 1703.00054}}].

\bibitem{Dutta:2015vwa}
B.~Dutta, R.~Mahapatra, L.~E. Strigari, and J.~W. Walker, {\it {Sensitivity to
  $Z$-prime and nonstandard neutrino interactions from ultralow threshold
  neutrino-nucleus coherent scattering}},  {\em Phys. Rev.} {\bf D93} (2016),
  no.~1 013015, [\href{http://www.arxiv.org/abs/1508.07981}{{\tt 1508.07981}}].

\bibitem{Denton:2018xmq}
P.~B. Denton, Y.~Farzan, and I.~M. Shoemaker, {\it {A Plan to Rule out Large
  Non-Standard Neutrino Interactions After COHERENT Data}},  {\em JHEP} {\bf
  07} (2018) 037, [\href{http://www.arxiv.org/abs/1804.03660}{{\tt
  1804.03660}}].

\bibitem{Lindner:2016wff}
M.~Lindner, W.~Rodejohann, and X.-J. Xu, {\it {Coherent Neutrino-Nucleus
  Scattering and new Neutrino Interactions}},  {\em JHEP} {\bf 03} (2017) 097,
  [\href{http://www.arxiv.org/abs/1612.04150}{{\tt 1612.04150}}].

\bibitem{Coloma:2017ncl}
P.~Coloma, M.~C. Gonzalez-Garcia, M.~Maltoni, and T.~Schwetz, {\it {COHERENT
  Enlightenment of the Neutrino Dark Side}},  {\em Phys. Rev.} {\bf D96}
  (2017), no.~11 115007, [\href{http://www.arxiv.org/abs/1708.02899}{{\tt
  1708.02899}}].

\bibitem{Liao:2017uzy}
J.~Liao and D.~Marfatia, {\it {COHERENT constraints on nonstandard neutrino
  interactions}},  {\em Phys. Lett.} {\bf B775} (2017) 54--57,
  [\href{http://www.arxiv.org/abs/1708.04255}{{\tt 1708.04255}}].

\bibitem{Kosmas:2017tsq}
D.~K. Papoulias and T.~S. Kosmas, {\it {COHERENT constraints to conventional
  and exotic neutrino physics}},  {\em Phys. Rev.} {\bf D97} (2018), no.~3
  033003, [\href{http://www.arxiv.org/abs/1711.09773}{{\tt 1711.09773}}].

\bibitem{Farzan:2018gtr}
Y.~Farzan, M.~Lindner, W.~Rodejohann, and X.-J. Xu, {\it {Probing neutrino
  coupling to a light scalar with coherent neutrino scattering}},  {\em JHEP}
  {\bf 05} (2018) 066, [\href{http://www.arxiv.org/abs/1802.05171}{{\tt
  1802.05171}}].

\bibitem{Abdullah:2018ykz}
M.~Abdullah, J.~B. Dent, B.~Dutta, G.~L. Kane, S.~Liao, and L.~E. Strigari,
  {\it {Coherent elastic neutrino nucleus scattering as a probe of a $Z'$
  through kinetic and mass mixing effects}},  {\em Phys. Rev.} {\bf D98}
  (2018), no.~1 015005, [\href{http://www.arxiv.org/abs/1803.01224}{{\tt
  1803.01224}}].

\bibitem{Billard:2018jnl}
J.~Billard, J.~Johnston, and B.~J. Kavanagh, {\it {Prospects for exploring New
  Physics in Coherent Elastic Neutrino-Nucleus Scattering}},
  \href{http://www.arxiv.org/abs/1805.01798}{{\tt 1805.01798}}.

\bibitem{nu-magnetic1}
A.~Dodd, E.~Papageorgiu, and S.~Ranfone, {\it The effect of a neutrino magnetic
  moment on nuclear excitation processes},  {\em Physics Letters B} {\bf 266}
  (1991), no.~3 434 -- 438.

\bibitem{nu-magnetic2}
T.~S. Kosmas, O.~G. Miranda, D.~K. Papoulias, M.~T\'ortola, and J.~W.~F. Valle,
  {\it Probing neutrino magnetic moments at the spallation neutron source
  facility},  {\em Phys. Rev. D} {\bf 92} (Jul, 2015) 013011.

\bibitem{Akhmedov:2018wlf}
E.~Akhmedov, G.~Arcadi, M.~Lindner, and S.~Vogl, {\it {Coherent scattering and
  macroscopic coherence: Implications for neutrino, dark matter and axion
  detection}},  \href{http://www.arxiv.org/abs/1806.10962}{{\tt 1806.10962}}.

\bibitem{CONUStalk}
W.~Maneschg, ``The status of conus.''
  \url{{https://doi.org/10.5281/zenodo.1286927}}, June, 2018.

\bibitem{Strauss:2017cuu}
R.~Strauss {\em et~al.}, {\it {The $\nu$-cleus experiment: A gram-scale
  fiducial-volume cryogenic detector for the first detection of coherent
  neutrino-nucleus scattering}},  {\em Eur. Phys. J.} {\bf C77} (2017) 506,
  [\href{http://www.arxiv.org/abs/1704.04320}{{\tt 1704.04320}}].

\bibitem{Aguilar-Arevalo:2016khx}
{\bf CONNIE} {\bf Collaboration}, A.~Aguilar-Arevalo {\em et~al.}, {\it {The
  CONNIE experiment}},  {\em J. Phys. Conf. Ser.} {\bf 761} (2016), no.~1
  012057, [\href{http://www.arxiv.org/abs/1608.01565}{{\tt 1608.01565}}].

\bibitem{Agnolet:2016zir}
{\bf MINER} {\bf Collaboration}, G.~Agnolet {\em et~al.}, {\it {Background
  Studies for the MINER Coherent Neutrino Scattering Reactor Experiment}},
  {\em Nucl. Instrum. Meth.} {\bf A853} (2017) 53--60,
  [\href{http://www.arxiv.org/abs/1609.02066}{{\tt 1609.02066}}].

\bibitem{Wong:2010zzc}
H.~T. Wong, {\it {Neutrino-nucleus coherent scattering and dark matter searches
  with sub-keV germanium detector}},  {\em Nucl. Phys.} {\bf A844} (2010)
  229C--233C.

\bibitem{Belov:2015ufh}
V.~Belov {\em et~al.}, {\it {The $\nu$GeN experiment at the Kalinin Nuclear
  Power Plant}},  {\em JINST} {\bf 10} (2015), no.~12 P12011.

\bibitem{Billard:2016giu}
J.~Billard {\em et~al.}, {\it {Coherent Neutrino Scattering with Low
  Temperature Bolometers at Chooz Reactor Complex}},  {\em J. Phys.} {\bf G44}
  (2017), no.~10 105101, [\href{http://www.arxiv.org/abs/1612.09035}{{\tt
  1612.09035}}].

\bibitem{Bertuzzo:2017lwt}
E.~Bertuzzo, C.~J. Caniu~Barros, and G.~Grilli~di Cortona, {\it {MeV Dark
  Matter: Model Independent Bounds}},  {\em JHEP} {\bf 09} (2017) 116,
  [\href{http://www.arxiv.org/abs/1707.00725}{{\tt 1707.00725}}].

\bibitem{Hochberg:2017wce}
Y.~Hochberg, Y.~Kahn, M.~Lisanti, K.~M. Zurek, A.~G. Grushin, R.~Ilan, S.~M.
  Griffin, Z.-F. Liu, S.~F. Weber, and J.~B. Neaton, {\it {Detection of sub-MeV
  Dark Matter with Three-Dimensional Dirac Materials}},  {\em Phys. Rev.} {\bf
  D97} (2018), no.~1 015004, [\href{http://www.arxiv.org/abs/1708.08929}{{\tt
  1708.08929}}].

\bibitem{Dolan:2017xbu}
M.~J. Dolan, F.~Kahlhoefer, and C.~McCabe, {\it {Directly detecting sub-GeV
  dark matter with electrons from nuclear scattering}},  {\em Phys. Rev. Lett.}
  {\bf 121} (2018), no.~10 101801,
  [\href{http://www.arxiv.org/abs/1711.09906}{{\tt 1711.09906}}].

\bibitem{Hufnagel:2017dgo}
M.~Hufnagel, K.~Schmidt-Hoberg, and S.~Wild, {\it {BBN constraints on MeV-scale
  dark sectors. Part I. Sterile decays}},  {\em JCAP} {\bf 1802} (2018) 044,
  [\href{http://www.arxiv.org/abs/1712.03972}{{\tt 1712.03972}}].

\bibitem{Dutra:2018gmv}
M.~Dutra, M.~Lindner, S.~Profumo, F.~S. Queiroz, W.~Rodejohann, and
  C.~Siqueira, {\it {MeV Dark Matter Complementarity and the Dark Photon
  Portal}},  {\em JCAP} {\bf 1803} (2018) 037,
  [\href{http://www.arxiv.org/abs/1801.05447}{{\tt 1801.05447}}].

\bibitem{Berlin:2018sjs}
A.~Berlin, D.~Hooper, G.~Krnjaic, and S.~D. McDermott, {\it {Severely
  Constraining Dark Matter Interpretations of the 21-cm Anomaly}},  {\em Phys.
  Rev. Lett.} {\bf 121} (2018), no.~1 011102,
  [\href{http://www.arxiv.org/abs/1803.02804}{{\tt 1803.02804}}].

\bibitem{Mertig:1990an}
R.~Mertig, M.~Bohm, and A.~Denner, {\it {FEYN CALC: Computer algebraic
  calculation of Feynman amplitudes}},  {\em Comput. Phys. Commun.} {\bf 64}
  (1991) 345--359.

\bibitem{Shtabovenko:2016sxi}
V.~Shtabovenko, R.~Mertig, and F.~Orellana, {\it {New Developments in FeynCalc
  9.0}},  {\em Comput. Phys. Commun.} {\bf 207} (2016) 432--444,
  [\href{http://www.arxiv.org/abs/1601.01167}{{\tt 1601.01167}}].

\bibitem{Coloma:2017egw}
P.~Coloma, P.~B. Denton, M.~C. Gonzalez-Garcia, M.~Maltoni, and T.~Schwetz,
  {\it {Curtailing the Dark Side in Non-Standard Neutrino Interactions}},  {\em
  JHEP} {\bf 04} (2017) 116, [\href{http://www.arxiv.org/abs/1701.04828}{{\tt
  1701.04828}}].

\bibitem{Huber:2011wv}
P.~Huber, {\it {On the determination of anti-neutrino spectra from nuclear
  reactors}},  {\em Phys. Rev.} {\bf C84} (2011) 024617,
  [\href{http://www.arxiv.org/abs/1106.0687}{{\tt 1106.0687}}]. [Erratum: Phys.
  Rev.C85,029901(2012)].

\bibitem{Mueller:2011nm}
T.~A. Mueller {\em et~al.}, {\it {Improved Predictions of Reactor Antineutrino
  Spectra}},  {\em Phys. Rev.} {\bf C83} (2011) 054615,
  [\href{http://www.arxiv.org/abs/1101.2663}{{\tt 1101.2663}}].

\bibitem{Kerman:2016jqp}
{\bf TEXONO} {\bf Collaboration}, S.~Kerman, V.~Sharma, M.~Deniz, H.~T. Wong,
  J.~W. Chen, H.~B. Li, S.~T. Lin, C.~P. Liu, and Q.~Yue, {\it {Coherency in
  Neutrino-Nucleus Elastic Scattering}},  {\em Phys. Rev.} {\bf D93} (2016),
  no.~11 113006, [\href{http://www.arxiv.org/abs/1603.08786}{{\tt
  1603.08786}}].

\bibitem{Kopeikin:2012zz}
V.~I. Kopeikin, {\it {Flux and spectrum of reactor antineutrinos}},  {\em Phys.
  Atom. Nucl.} {\bf 75} (2012) 143--152. [Yad. Fiz.75N2,165(2012)].

\bibitem{Buck:2015clx}
C.~Buck, A.~P. Collin, J.~Haser, and M.~Lindner, {\it {Investigating the
  Spectral Anomaly with Different Reactor Antineutrino Experiments}},  {\em
  Phys. Lett.} {\bf B765} (2017) 159--162,
  [\href{http://www.arxiv.org/abs/1512.06656}{{\tt 1512.06656}}].

\bibitem{Giunti:2016elf}
C.~Giunti, {\it {Precise determination of the $^{235}$U reactor antineutrino
  cross section per fission}},  {\em Phys. Lett.} {\bf B764} (2017) 145--149,
  [\href{http://www.arxiv.org/abs/1608.04096}{{\tt 1608.04096}}].

\bibitem{Huber:2016xis}
P.~Huber, {\it {NEOS Data and the Origin of the 5 MeV Bump in the Reactor
  Antineutrino Spectrum}},  {\em Phys. Rev. Lett.} {\bf 118} (2017), no.~4
  042502, [\href{http://www.arxiv.org/abs/1609.03910}{{\tt 1609.03910}}].

\bibitem{COH-talk}
G.~Rich, ``The coherent collaboration and the first observation of coherent
  elastic neutrino-nucleus scattering.''
  \url{{https://doi.org/10.5281/zenodo.1286967}}, June, 2018.

\bibitem{Chang:2018rso}
J.~H. Chang, R.~Essig, and S.~D. McDermott, {\it {Supernova 1987A Constraints
  on Sub-GeV Dark Sectors, Millicharged Particles, the QCD Axion, and an
  Axion-like Particle}},  {\em JHEP} {\bf 09} (2018) 051,
  [\href{http://www.arxiv.org/abs/1803.00993}{{\tt 1803.00993}}].

\bibitem{Davidson:2000hf}
S.~Davidson, S.~Hannestad, and G.~Raffelt, {\it {Updated bounds on millicharged
  particles}},  {\em JHEP} {\bf 05} (2000) 003,
  [\href{http://www.arxiv.org/abs/hep-ph/0001179}{{\tt hep-ph/0001179}}].

\bibitem{Foot:2015sia}
R.~Foot, {\it {Dissipative dark matter explains rotation curves}},  {\em Phys.
  Rev.} {\bf D91} (2015), no.~12 123543,
  [\href{http://www.arxiv.org/abs/1502.07817}{{\tt 1502.07817}}].

\bibitem{Wolfenstein}
L.~Wolfenstein, {\it {Neutrino Oscillations in Matter}},  {\em Phys.Rev.} {\bf
  D17} (1978) 2369--2374.

\bibitem{Mikheev:1986gs}
S.~P. Mikheev and A.~{\relax Yu}. Smirnov, {\it {Resonance Amplification of
  Oscillations in Matter and Spectroscopy of Solar Neutrinos}},  {\em Sov. J.
  Nucl. Phys.} {\bf 42} (1985) 913--917. [Yad. Fiz.42,1441(1985)].

\bibitem{Mikheev:1986wj}
S.~P. Mikheev and A.~{\relax Yu}. Smirnov, {\it {Resonant amplification of
  neutrino oscillations in matter and solar neutrino spectroscopy}},  {\em
  Nuovo Cim.} {\bf C9} (1986) 17--26.

\bibitem{Arguelles:2016uwb}
C.~A. Argüelles, V.~Brdar, and J.~Kopp, {\it {Production of keV Sterile
  Neutrinos in Supernovae: New Constraints and Gamma Ray Observables}},
  \href{http://www.arxiv.org/abs/1605.00654}{{\tt 1605.00654}}.

\bibitem{Goran}
R.~N. Mohapatra and G.~Senjanovi\ifmmode~\acute{c}\else \'{c}\fi{}, {\it
  Neutrino mass and spontaneous parity nonconservation},  {\em Phys. Rev.
  Lett.} {\bf 44} (Apr, 1980) 912--915.

\bibitem{Minkowski}
P.~Minkowski, {\it $\mu\to e\gamma$ at a rate of one out of 109 muon decays?},
  {\em Physics Letters B} {\bf 67} (1977), no.~4 421 -- 428.

\bibitem{GellMann:1980vs}
M.~Gell-Mann, P.~Ramond, and R.~Slansky, {\it {Complex Spinors and Unified
  Theories}},  {\em Conf. Proc.} {\bf C790927} (1979) 315--321,
  [\href{http://www.arxiv.org/abs/1306.4669}{{\tt 1306.4669}}].

\bibitem{Yanagida:1979as}
T.~Yanagida, {\it {Horizontal Symmetry and Masses of Neutrinos}},  {\em Conf.
  Proc.} {\bf C7902131} (1979) 95--99.

\bibitem{Atre:2009rg}
A.~Atre, T.~Han, S.~Pascoli, and B.~Zhang, {\it {The Search for Heavy Majorana
  Neutrinos}},  {\em JHEP} {\bf 05} (2009) 030,
  [\href{http://www.arxiv.org/abs/0901.3589}{{\tt 0901.3589}}].

\bibitem{deGouvea:2015euy}
A.~de~Gouvêa and A.~Kobach, {\it {Global Constraints on a Heavy Neutrino}},
  {\em Phys. Rev.} {\bf D93} (2016), no.~3 033005,
  [\href{http://www.arxiv.org/abs/1511.00683}{{\tt 1511.00683}}].

\bibitem{Deppisch:2015qwa}
F.~F. Deppisch, P.~S. Bhupal~Dev, and A.~Pilaftsis, {\it {Neutrinos and
  Collider Physics}},  {\em New J. Phys.} {\bf 17} (2015), no.~7 075019,
  [\href{http://www.arxiv.org/abs/1502.06541}{{\tt 1502.06541}}].

\bibitem{Vincent:2014rja}
A.~C. Vincent, E.~F. Martinez, P.~Hernández, M.~Lattanzi, and O.~Mena, {\it
  {Revisiting cosmological bounds on sterile neutrinos}},  {\em JCAP} {\bf
  1504} (2015), no.~04 006, [\href{http://www.arxiv.org/abs/1408.1956}{{\tt
  1408.1956}}].

\bibitem{Bernardi:1985ny}
G.~Bernardi {\em et~al.}, {\it {Search for Neutrino Decay}},  {\em Phys. Lett.}
  {\bf 166B} (1986) 479--483.

\bibitem{Acciarri:2016crz}
{\bf DUNE} {\bf Collaboration}, R.~Acciarri {\em et~al.}, {\it {Long-Baseline
  Neutrino Facility (LBNF) and Deep Underground Neutrino Experiment (DUNE)}},
  \href{http://www.arxiv.org/abs/1601.05471}{{\tt 1601.05471}}.

\bibitem{Lantwin:2017xtc}
O.~Lantwin, {\it {Search for new physics with the SHiP experiment at CERN}},
  {\em PoS} {\bf EPS-HEP2017} (2017) 304,
  [\href{http://www.arxiv.org/abs/1710.03277}{{\tt 1710.03277}}].

\bibitem{Feng:2017uoz}
J.~L. Feng, I.~Galon, F.~Kling, and S.~Trojanowski, {\it {ForwArd Search
  ExpeRiment at the LHC}},  {\em Phys. Rev.} {\bf D97} (2018), no.~3 035001,
  [\href{http://www.arxiv.org/abs/1708.09389}{{\tt 1708.09389}}].

\bibitem{Feng:2017vli}
J.~L. Feng, I.~Galon, F.~Kling, and S.~Trojanowski, {\it {Dark Higgs bosons at
  the ForwArd Search ExpeRiment}},  {\em Phys. Rev.} {\bf D97} (2018), no.~5
  055034, [\href{http://www.arxiv.org/abs/1710.09387}{{\tt 1710.09387}}].

\bibitem{Kling:2018wct}
F.~Kling and S.~Trojanowski, {\it {Heavy Neutral Leptons at FASER}},  {\em
  Phys. Rev.} {\bf D97} (2018), no.~9 095016,
  [\href{http://www.arxiv.org/abs/1801.08947}{{\tt 1801.08947}}].

\bibitem{CortinaGil:2017mqf}
{\bf NA62} {\bf Collaboration}, E.~Cortina~Gil {\em et~al.}, {\it {Search for
  heavy neutral lepton production in $K^+$ decays}},  {\em Phys. Lett.} {\bf
  B778} (2018) 137--145, [\href{http://www.arxiv.org/abs/1712.00297}{{\tt
  1712.00297}}].

\bibitem{Curtin:2018mvb}
D.~Curtin {\em et~al.}, {\it {Long-Lived Particles at the Energy Frontier: The
  MATHUSLA Physics Case}},  \href{http://www.arxiv.org/abs/1806.07396}{{\tt
  1806.07396}}.

\bibitem{Xing:2011}
Z.~Xing and S.~Zhou, {\em Neutrinos in Particle Physics, Astronomy and
  Cosmology}.
\newblock Advanced Topics in Science and Technology in China. Springer Berlin
  Heidelberg, 2011.

\bibitem{deNiverville:2015mwa}
P.~deNiverville, M.~Pospelov, and A.~Ritz, {\it {Light new physics in coherent
  neutrino-nucleus scattering experiments}},  {\em Phys. Rev.} {\bf D92}
  (2015), no.~9 095005, [\href{http://www.arxiv.org/abs/1505.07805}{{\tt
  1505.07805}}].

\bibitem{Ge:2017mcq}
S.-F. Ge and I.~M. Shoemaker, {\it {Constraining Photon Portal Dark Matter with
  Texono and Coherent Data}},  \href{http://www.arxiv.org/abs/1710.10889}{{\tt
  1710.10889}}.

\bibitem{Garny:2017rxs}
M.~Garny, J.~Heisig, B.~Lülf, and S.~Vogl, {\it {Coannihilation without
  chemical equilibrium}},  {\em Phys. Rev.} {\bf D96} (2017), no.~10 103521,
  [\href{http://www.arxiv.org/abs/1705.09292}{{\tt 1705.09292}}].

\bibitem{Ellis:2015vaa}
J.~Ellis, F.~Luo, and K.~A. Olive, {\it {Gluino Coannihilation Revisited}},
  {\em JHEP} {\bf 09} (2015) 127,
  [\href{http://www.arxiv.org/abs/1503.07142}{{\tt 1503.07142}}].

\bibitem{Gondolo:1990dk}
P.~Gondolo and G.~Gelmini, {\it {Cosmic abundances of stable particles:
  Improved analysis}},  {\em Nucl. Phys.} {\bf B360} (1991) 145--179.

\bibitem{Aghanim:2018eyx}
{\bf Planck} {\bf Collaboration}, N.~Aghanim {\em et~al.}, {\it {Planck 2018
  results. VI. Cosmological parameters}},
  \href{http://www.arxiv.org/abs/1807.06209}{{\tt 1807.06209}}.

\bibitem{Hall:2009bx}
L.~J. Hall, K.~Jedamzik, J.~March-Russell, and S.~M. West, {\it {Freeze-In
  Production of FIMP Dark Matter}},  {\em JHEP} {\bf 03} (2010) 080,
  [\href{http://www.arxiv.org/abs/0911.1120}{{\tt 0911.1120}}].

\bibitem{Molinaro:2014lfa}
E.~Molinaro, C.~E. Yaguna, and O.~Zapata, {\it {FIMP realization of the
  scotogenic model}},  {\em JCAP} {\bf 1407} (2014) 015,
  [\href{http://www.arxiv.org/abs/1405.1259}{{\tt 1405.1259}}].

\bibitem{Baumholzer:2018sfb}
S.~Baumholzer, V.~Brdar, and P.~Schwaller, {\it {The New $\nu$MSM
  ($\nu\nu$MSM): Radiative Neutrino Masses, keV-Scale Dark Matter and Viable
  Leptogenesis with sub-TeV New Physics}},  {\em JHEP} {\bf 08} (2018) 067,
  [\href{http://www.arxiv.org/abs/1806.06864}{{\tt 1806.06864}}].

\bibitem{Essig:2015cda}
R.~Essig, M.~Fernandez-Serra, J.~Mardon, A.~Soto, T.~Volansky, and T.-T. Yu,
  {\it {Direct Detection of sub-GeV Dark Matter with Semiconductor Targets}},
  {\em JHEP} {\bf 05} (2016) 046,
  [\href{http://www.arxiv.org/abs/1509.01598}{{\tt 1509.01598}}].

\bibitem{Slatyer:2015jla}
T.~R. Slatyer, {\it {Indirect dark matter signatures in the cosmic dark ages.
  I. Generalizing the bound on s-wave dark matter annihilation from Planck
  results}},  {\em Phys. Rev.} {\bf D93} (2016), no.~2 023527,
  [\href{http://www.arxiv.org/abs/1506.03811}{{\tt 1506.03811}}].

\end{thebibliography}\endgroup

\end{document}